\documentclass{article} 

\usepackage{amsmath}   
\usepackage{amsfonts} 
\usepackage{amsthm}
\usepackage{epsfig} 
\usepackage{subfigure}

\usepackage{graphicx} 

\usepackage{tikz}
\usetikzlibrary{shapes.gates.logic.US,trees,positioning,arrows}

\newtheorem{theorem}{Theorem}

\newcommand{\mx}[1]{\boldsymbol{#1}}
\newcommand{\vect}[1]{\boldsymbol{#1}}

\newcommand{\idvt}{\,{\rm I\kern-0.55em 1 }}
\newcommand{\idvts}{\,{\rm I\kern-0.25em 1 }}
\newcommand{\exitv}{\vect{\psi}}
\newcommand{\exitvp}{\vect{\psi'}}

\newcommand{\iniv}{\vect{\pi}}
\newcommand{\inivp}{\vect{\pi'}}

\newcommand{\statv}{\vect{\phi}}
\newcommand{\addv}{\vect{\xi}}
\newcommand{\intv}{\vect{\nu}}
\newcommand{\pathv}{\vect{\upsilon}}

\DeclareMathOperator{\diag}{diag}

\begin{document}
\title{On the Representation of Correlated Exponential Distributions by Phase Type Distributions}
\author{Peter Buchholz
Informatik IV, TU Dortmund\\
D-44221 Dortmund, Germany\\
{\small \textit{peter.buchholz@cs.tu-dortmund.de}}
}
         
\maketitle         
\begin{abstract}
In this paper we present  results for bivariate exponential distributions which are represented by phase type distributions. The paper extends results from previous publications \cite{BlNi10,HeZV12} on this topic by introducing new representations that require a smaller number of phases to reach some correlation coefficient and introduces different ways to describe correlation between exponentially distributed random variables. Furthermore, it is shown how \textit{Markovian Arrival Processes} (MAPs) with exponential marginal distribution can be generated from the phase type representations of exponential distributions and how the results for exponential distributions can be applied to define correlated hyperexponential or Erlang distributions. As application examples we analyze two queueing models with correlated inter-arrival and service times. \bigskip           
\end{abstract}

\section{Introduction}

\textit{Phase type distributions} (PHDs) are used to describe non-exponential distributions in stochastic models that can be mapped on \textit{Continuous Time Markov Chains} (CTMCs) and solved by simulation or,  preferably, by numerical techniques \cite{Neut81,Stew09}. A large number of papers on PHDs, their properties, the estimation of parameters, and their application in stochastic modeling exists \cite{ArGH10,BKF14}.  PHDs describe uncorrelated event streams whereas their extension \textit{Markovian Arrival Processes} (MAPs) are applied to model autocorrelated sequences. In many stochastic models correlation does not only occur between the events of one stream, instead the times of different events are correlated. E.g., inter-arrival times and service times in a queuing system are correlated \cite{CiBS21}, failures times of components are correlated \cite{GMSR18} or processing times of parallel jobs are correlated \cite{LiPF14}. In these cases, correlation between different PHDs has to be modeled. To describe the required correlation, PHDs that run in parallel or sequentially have to be defined to realize correlation of processing times of parallel jobs or of sequential steps of one job. 

In this paper, we consider mainly correlated exponential distributions which are the base for PHDs and MAPs. Results from two previous papers on bivariate exponential distributions \cite{BlNi10,HeZV12} are extended. New representations for bivariate or multi-variate exponential distributions described by PHDs are defined that require less states to represent a given coefficient of correlation. Furthermore, some applications of correlated exponential distributions are presented. Phase type representations of the exponential distribution can be furthermore used as building blocks for general PHDs and for MAPs to model autocorrelated sequences of random variables. Both aspects will also be briefly considered.

The paper is structured as follows. In the next section PHDs and phase type representations of exponential distributions are introduced. Section~\ref{sec:related} reviews related work. Then,  a stepwise approach to generate phase type representations of exponential distributions with an increasing number of phases is defined. The following two sections describe how a maximal and minimal coefficient of correlation can be achieved by phase type representations of exponential distributions. For the positive correlation, results from \cite{BlNi10} are improved by  finding a representation with less phases to reach a given coefficient of correlation. For negative correlation, it is shown that the PHD from \cite{BlNi10} observes some local optimality property but is not globally optimal.  Then, we consider MAPs with exponential marginal distribution in Section~\ref{sec:map} and the extension from exponential to hyperexponential and Erlang distributions in Section~\ref{sec:general}. Finally, two queueing models are presented as application examples. 

\section{Basic Model}

We begin with a short introduction of PHDs before we consider the representation of exponential distributions by PHDs. 
A PHD is described by an initial distribution vector $\iniv$ and a sub-generator $\mx{D}$ with only transient states \cite{BKF14,Neut81}. Consequently, the real part of all eigenvalues of matrix $\mx{D}$ is negative implying that the matrix is non-singular and the inverse matrix is non-positive. Furthermore, we assume $\iniv\idvt = 1$. PHDs can be interpreted as absorbing Markov chains with the following initial vector and generator matrix.
\begin{equation}
\left(\iniv, 0\right) \mbox{ and } \left( \begin{array}{cc} \mx{D} & \mx{d}\\\mx{0} & 0\end{array}\right)
\end{equation}
where $\mx{d} = -\mx{D}\idvt$ and $\idvt$ is a column vectors of all $1$s. Let $n$ be the number of phases of the PHD, then the absorbing Markov chain contains $n+1$ states, $n$ transient states and one absorbing state. 
 
For some PHD $(\iniv, \mx{D})$, matrix $\mx{M} = -\mx{D}^{-1} \ge \mx{0}$ exists, $\mx{M}(i,j)$ is the mean time the PHD stays in state $j$ before absorption if the current state is $i$ and $\mx{m} = \mx{M}\idvt$ is a vector containing the first moments of the time to absorption conditioned on the initial state. Let $\exitv$ be the vector of exit probabilities from the states, containing the probabilities that the PHD is left from a specific state. The vector can be computed as follows \cite{KeS69}.
\begin{equation}
\mx{B} = \mx{M}\diag\left(\mx{d}\right) \mbox{ and } \exitv = \iniv\mx{B} . 
\end{equation}
Another quantity which is of interest is the expected absorption time conditioned on the exit state (i.e, the last transient state before absorption). Let $\mx{a}$ be the vector containing the conditional absorption times. The vector can be computed as 
\begin{equation}
\label{eq:exit_time}
\mx{a}(i) =\frac{1}{\exitv(i)} \iniv\mx{M} \mx{B}(:,i) \mbox{ for } \exitv(i) > 0 
\end{equation}
where $\mx{B}(:,i)$ is the $i$th column of matrix $\mx{B}$. This equation can be derived from \cite{KeS69}. 

We consider  \textit{Acyclic Phase Type Distributions} (APHDs)  which are described by the initial distribution $\iniv$ and a matrix $\mx{D}$ that can be permuted to an upper triangular matrix.  In contrast to general PHDs, APHDs can be transformed to some canonical form by equivalence transformations. Different canonical forms for  APHDs and the corresponding transformation rules are proposed in \cite{Cuma82}. We consider in the following two canonical forms. The first is given by
\begin{equation}
\label{eq:first_canonical}
\iniv = \left(\pi_1,\ldots,\pi_n\right), \
\mx{D} = \left(\begin{array}{ccccc}
-\mu_1 & \mu_1 & 0 & \cdots & 0\\
0 & \ddots & \ddots & \ddots  & \vdots\\
\vdots & \vdots & \ddots & \ddots  & \vdots\\
\vdots & \vdots &\vdots & -\mu_{n-1} & \mu_{n-1}\\
0 & \cdots & \cdots & 0 & -\mu_n
\end{array}\right)
\end{equation}
with $\mu_1 \le \mu_2 \le \ldots \le \mu_n$. 
The second canonical form equals 
\begin{equation}
\label{eq:second_canonical}
\iniv = \left(1,0,\ldots,0\right), \
\mx{D} = \left(\begin{array}{ccccc}
-\mu_1 & \mu_{1,2} & 0 & \cdots & 0\\
0 & \ddots & \ddots & \ddots & \vdots\\
\vdots & \vdots & \ddots & \ddots & \vdots\\
\vdots & \vdots & \vdots &  -\mu_{n-1} & \mu_{n-1,n}\\
0 & \cdots & \cdots & 0 & -\mu_n
\end{array}\right)
\end{equation}
where $\mu_1 \ge \mu_2 \ge \ldots \ge \mu_n $. The Laplace transform of the distribution described by the first or second canonical representation is given by
\begin{equation}
L_1(s) = \sum_{i=1}^n \pi_i \prod_{j=i}^n \frac{\mu_j}{\mu_j +s} \mbox{ and }
L_2(s) = \sum_{i=1}^n\left(1-\frac{\mu_{i,i+1}}{\mu_i}\right) \left( \prod_{j=1}^{i-1} \frac{\mu_{j,j+1}}{\mu_{j}}\right)\left(\prod_{k=1}^i\frac{\mu_k}{\mu_k+s}\right) ,
\end{equation}
where $\mu_{n,n+1}=0$. Our goal is to represent the exponential distribution by APHDs to model correlated exponential distributions. To represent some exponential distribution with rate $\lambda$ by an APHD in the first canonical form $L_1(s) = \frac{\lambda}{\lambda+s}$ has to hold, to represent it in the second canonical form $L_2(s) = \frac{\lambda}{\lambda+s}$ has to hold. We denote a representation as normalized if $\lambda=1$ which implies $E(X)=1$. By multiplication of matrix $\mx{D}$ with $\lambda^{-1}>0$ we obtain a normalized representation.  Thus, it is sufficient to consider only normalized representations in the sequel. 

If we assume that a PHD is immediately started again with vector $\iniv$ after an absorption, we obtain the stationary vector at random times $\statv$ as the solution of the following set of equations.
\begin{equation}
\statv \left( \mx{D} + \mx{d}\iniv\right) = \mx{0} \mbox{ and } \statv\idvt = 1 .
\end{equation}

For some random variable $X$ we denote by $(\iniv_X,\mx{D}_X)$ a PHD representation. If two PHDs $(\iniv_X,\mx{D}_X)$ and $(\inivp_X,\mx{D'}_X)$ are different representations for the same random variable we use the notation $(\iniv_X,\mx{D}_X) \sim (\inivp_X,\mx{D'}_X)$. For $X$ with representation $(\iniv_X,\mx{D}_X)$, the relations $E(X) = \left(\statv_X\mx{d}_X\right)^{-1} = \iniv_X\mx{m}_X$ hold. 

For two random variables $X$ and $Y$,  the correlation and coefficient of correlation are defined as
\begin{equation}
C_{X,Y} = E(XY)-E(X)E(Y) \mbox{ and } \rho_{X,Y} = \frac{C_{X,Y}}{\sigma_X\sigma_Y}
\end{equation}
where $\sigma_X,\sigma_Y$ are the standard deviations. For the exponential distribution $\sigma_X = E(X)$ such that $\rho_{X,Y} = \frac{E(XY)}{E(X)E(Y)} - 1$ which implies for two normalized representations $\rho_{X,Y} = E(XY) - 1$. Let $Y$ and $X$ be two exponentially distributed random variables. We assume that $(\iniv_X,\mx{D}_X)$ and $(\iniv_Y,\mx{D}_Y)$ are APHDs representing normalized random variables. By multiplying the matrices $\mx{D}_X$ and $\mx{D}_Y$ with $\lambda_X$ and $\lambda_Y$, respectively, the rates of the distributions can be shifted.

We describe two different possibilities to combine PHDs. The bivariate representation used in \cite{BlNi10,HeZV12} can be interpreted as a sequential composition of PHDs. Let $(\iniv_X,\mx{D}_X)$ and  $(\iniv_Y,\mx{D}_Y)$ be two PHDs of order $n_X$ and $n_Y$ describing random variables $X$ and $Y$, then the following absorbing CTMC defines the sequential combination. 
\begin{equation}
\label{eq:expansion_out_in}
\left(\iniv_X,\mx{0}, 0\right), \ \left(\begin{array}{ccc}\mx{D}_X & \diag(\mx{d}_X)\mx{\Psi}_{X,Y} & \mx{0}\\ \mx{0} & 
\mx{D}_Y & \mx{d}_Y\\\mx{0} & \mx{0} & 0\end{array}\right)
\end{equation}
where  $\mx{\Psi}$ is a non-negative $n_X \times n_Y$ matrix with $\mx{\Psi}_{X,Y}\idvt = \idvt$ and $\exitv_X\mx{\Psi}_{X,Y} = \iniv_Y$.
The time of the transition from the first block into the second block determines the value of the first random variable, the exit state from where the first block is left defines the initial distribution for the second random variable, and the time of the transition into the absorbing state determines the sum of both random variables and thus the value of the second random variable.  The coefficient of correlation of $X$ and $Y$ is then given by
\begin{equation}
\rho_{X,Y} = \frac{\left(\sum\limits_{i=1}^{n_X}\exitv_X(i)\mx{a}_X(i)\sum\limits_{j=1}^{n_Y}\mx{\Psi}_{X,Y}(i,j)\mx{m}_Y(j)\right) - E(X)E(Y)}{\sigma_X\sigma_Y} 
\end{equation}

The second composition is a parallel composition. In this case, two (or more) PHDs are started jointly. The combined PHDs built again a PHD described by the following absorbing CTMC.
\begin{equation}
\label{eq:expansion_parallel}
\left(\iniv_{X,Y}, \mx{0}, \mx{0}, 0\right) , \ 
\left(\begin{array}{cccc}
\mx{D}_X \oplus \mx{D}_Y & \mx{d}_X \otimes \mx{I}_{n_Y} & \mx{I}_{n_X} \otimes \mx{d}_Y & \mx{0} \\
\mx{0} & \mx{D}_Y & \mx{0} & \mx{d}_Y \\
\mx{0} & \mx{0}& \mx{D}_X & \mx{d}_X \\
\mx{0} & \mx{0} & \mx{0} & 0 
\end{array}\right)
\end{equation}
$\iniv_{X,Y}$ is a probability distribution which observes $\iniv_{X,Y} \left(\mx{I}_{n_X} \otimes \idvt_{n_Y}\right) = \iniv_X$ and  $\iniv_{X,Y} \left(\idvt_{n_X} \otimes \mx{I}_{n_Y}\right) = \iniv_Y$  where $\mx{I}_n$ is the $n \times n$ identity matrix and $\idvt_n$ a column vector of $1$s of length $n$. The time of a transition from the first into the second block describes the value of $X$ if it is smaller than the value of $Y$, a transition from the first into the third block describes the value of $Y$ if it is smaller than $X$ and a transition into the absorbing state defines the maximum of both random variables. The coefficient of correlation equals 
\begin{equation}
\rho_{X,Y} = \frac{\left(\sum\limits_{i=1}^{n_X}\sum\limits_{j=1}^{n_Y}\iniv_{X,Y}((i-1)*n_Y+j)\mx{m}_X(i)\mx{m}_Y(j)\right) - E(X)E(Y)}{\sigma_X\sigma_Y} 
\end{equation} 
The coefficient of correlation depends on vector $\iniv_{X,Y}$ and reachable coefficients of correlation depend on $\iniv_X$, $\iniv_Y$, $\mx{m}_X$ and $\mx{m}_Y$. It is straightforward to extend the parallel composition  to more than two PHDs but the state space grows with product of the dimensions of the PHDs  \cite{BBS20}. 

Obviously, the coefficients of correlation that can be reached for two random variables $X$ and $Y$ that are composed sequentially or in parallel depend on the choice of matrix $\mx{\Psi}_{X,Y}$ and vector $\iniv_{X,Y}$, respectively. Additionally, the value depends also on the representation of the random variables by PHDs. Finding an appropriate representation to reach a given coefficient of correlation with a small dimension of the PHDs is a challenging problem that will be investigated in this paper.

\section{Related Work}
\label{sec:related}

As already mentioned, an enormous number of papers on PHDs, MAPs and stochastic models based on these models exist. For overviews including different application areas we refer to \cite{Aa95,ArGH10,BKF14}. Most important for our work are results on bivariate PHDs and in particular on bivariate exponential distributions which will be briefly reviewed here. 

The basic approach to generate multivariate PHDs is based on \cite{Ku89}. The idea is to define a PHD $(\iniv,\mx{D})$ of order $n$ and an $n \times r$ reward matrix $\mx{R}$ where $\mx{R}(:,j)$ is column $j$ of $\mx{R}$. Then the random variable $Y_j = \int_0^\infty \iniv_t\mx{R}(:,j)dt$ with $\iniv_t = \iniv e^{\mx{D}t}$ is phase type distributed and the different random variables $Y_j$ can be correlated or uncorrelated based on the choice of the basic PHD and matrix $\mx{R}$. The phase type model of \cite{Ku89} is hard to analyze and parameters are hard to fit according to predefined characteristics to be matched by the random variables. It is possible to generate realization of the random variables from the representation which may be used in simulation  but the representation cannot be used in Markov models solved numerically. 

Based on the model for multivariate PHD, \cite{BlNi10} developed a model of bivariate exponential PHDs using the representation (\ref{eq:expansion_out_in}).  To generate correlated exponential distributions, matrix $\mx{D}_X$ is of the second canonical form and $\mx{D}_Y$ of the first canonical form. Furthermore, $n_X = n_Y$ and $\mu_i = i$ ($i=1,\ldots,n_i$). In this case the minimal coefficient of correlation is reached for $\mx{\Psi}_{X,Y}=\mx{I}$ and the maximal coefficient of correlation is achieved for matrix $\mx{\Psi}_{X,Y}(i,j) = 1$ if $j=n_X-i+1$ and $0$ otherwise. For positive correlation, the maximal coefficient of correlation for order $n$ PHDs equals then $\rho^{+(n)} = 1 - \frac{1}{n}\sum_{i=1}^n\frac{1}{i}$ and the minimal coefficient of correlation equals $1-\sum_{i=1}^n\frac{1}{i^2}$. Thus, $\lim_{n\rightarrow \infty}\rho^{+(n)} = 1$ and $\lim_{n\rightarrow \infty}\rho^{-(n)} = 1 - \frac{\pi^2}{6}$, the minimal correlation coefficient for bivariate exponential distributions. 

To generate correlated exponential distributions with some coefficient of correlation $\rho > 0$, first the smallest $n$ such that $\rho^{+(n)} \ge \rho$ has been found and then $\mx{\Psi}_{X,Y} = \frac{\rho}{\rho^{+(n)}} \mx{I}_n + \left(1 - \frac{\rho}{\rho^{+(n)}}\right) \idvt\iniv_Y$ is selected resulting in the required coefficient of correlation. The generation of an appropriate matrix $\mx{\Psi}_{X,Y}$ for negative correlation works similarly.  The same idea can, of course, be applied to generate initial vectors for correlated PHDs composed in parallel. Let $(\iniv_X,\mx{D}_X)$ and $(\iniv_Y,\mx{D}_Y)$ be two identical PHDs of order $n$, then $\rho^{+(n)}$ is reached with initial vector $\iniv_{X,Y}(i,j) = 1$ for $i=j$ and $0$ otherwise. Correlation coefficient $\rho$ ($\ge \rho^{+(n)})$ can be obtained with initial vector $ \frac{\rho}{\rho^{+(n)}} \iniv_{X,Y} + (1-  \frac{\rho}{\rho^{+(n)}}) (\iniv_X \otimes \iniv_Y)$. Again, negative correlations can be achieved similarly.  

The results of \cite{BlNi10} are extended in \cite{HeZV12}.  The phase type representations used in \cite{BlNi10} representing exponential distributions according to (\ref{eq:expansion_out_in}) have the following properties.
\begin{itemize}
\item[i)] $\mx{D}_X\idvt = -\idvt$,
\item[ii)] $\idvt^T\mx{D}_Y=-\idvt^T$,
\item[iii)] $\idvt^T\mx{D}_X \le \mx{0}$,
\item[iv)] $\idvt^T\mx{\Psi}_{X,Y} = \idvt^T$.
\end{itemize}
\cite{HeZV12} shows that among all phase type representations of a fixed order $n$ that observe i)-iv), the representations proposed in \cite{BlNi10} have the smallest, respectively, largest coefficient of correlation. However, the properties are not mandatory to represent a normalized exponential distribution by a PHD or an APHD. Even if the canonical representations are used as phase type representations, not all properties hold. Property i) holds for the second canonical form (\ref{eq:second_canonical}), whereas property ii) does not necessarily hold for the first canonical form (\ref{eq:first_canonical}) because $1 < \mu_i \le \mu_{i+1}$ ($1 < i < n$) and not $\mu_{i+1} = \mu_i +1$ is required. Interestingly, the transformation of the first canonical form into the second canonical form via Theorem~\ref{th:transform} results in a phase type representation where property i) holds (see Theorem~\ref{th:identity}). This observation indicates that the conditions are not symmetric. Property iii) holds if $\mu_{i+1} \le \mu_{i}$ is used as it is necessary for the second canonical form but it is not necessary for a phase type representation. Property iv) depends on the solution of the LP (\ref{eq:lp_in_out}) which determines a matrix $\mx{\Psi}_{X,Y}$ that minimizes or maximizes the correlation coefficient for given representations of the PHDs. For negative correlation and representations with identical number of states  in the first and second canonical form $\mx{\Psi}_{X,Y}=\mx{I}$ is optimal and observes iv) only if $\exitv_X(i) = \iniv_Y(n-i+1)$ for all $i=1,\ldots n$. For positive correlation the optimal matrix $\mx{\Psi}_{X,Y}$ observes property iv).

Additionally, \cite{HeZV12} extends the result to phase type representations of different dimensions, i.e., $n_X \neq n_Y$. In this case the fourth property is substituted by
\begin{itemize}
\item[iv')] $\idvt^T_{n_X}\mx{\Psi}_{X,Y}=\frac{n_X}{n_Y}\idvt^T_{n_Y}$ .
\end{itemize}
Then, explicit matrices $\mx{\Psi}_{X,Y}$ are given to reach a minimal or maximal coefficient of correlation.  These matrices are the solutions of the LP (\ref{eq:lp_in_out}) in the cases where properties i)-iii) hold, but they do not necessarily hold if ii) does not hold. 

\section{Consecutive Generation of APHDs}
\label{sec:consecutive}

To represent an exponential distribution by an APHD in canonical form, the Laplace transforms of both have to coincide. The following theorem shows that this determines the rate of the first phase of an APHD in the first canonical form. 

\begin{theorem}
\label{th:first_phase}
If a normalized exponential distribution is represented by an APHD in canonical form (\ref{eq:first_canonical}), then $\mu_1 = 1$ has to hold. 
\end{theorem}
 
The proof of this and all other theorems can be found in the appendix. 
The proof of the theorem shows that $\iniv_t(i) = \iniv(i)e^{-t}$ holds for all states of the APHD in the first canonical form which represents an exponential distribution. This also implies that $\statv=\iniv$ holds in this case. 

Obviously, the coefficient of correlation of two exponential distributions composed as in (\ref{eq:expansion_out_in}) depends on matrix $\mx{\Psi}_{X,Y}$ which can be computed from a system of linear equations , if both PHDs and $\rho_{X,Y}$ are known. For given phase type representations the minmal or maximal coefficient of correlation can be computed from the following linear program. 
\begin{equation}
\label{eq:lp_in_out}
\begin{array}{ll}
\rho^\pm_{(\iniv_X,\mx{D}_X)(\iniv_Y,\mx{D}_Y)} = \min/\max \sum\limits_{i=1}^{n_X}\sum\limits_{j=1}^{n_Y} \mx{\Psi}_{X,Y}(i,j) \exitv_X(i)\vect{a}_X(i)\vect{m}_Y(j)-1\\
\mbox{s.t. } \mx{\psi}_X\mx{\Psi}_{X,Y} = \iniv_Y, \mx{\Psi}_{X,Y} \ge \mx{0}  \mbox{ and } \mx{\Psi}_{X,Y} \idvt = \idvt . 
\end{array}
\end{equation}

A similar LP can be derived for the composition (\ref{eq:expansion_parallel}) to compute initial vector $\iniv_{X,Y}$. 
\begin{equation}
\label{eq:lp_input}
\begin{array}{ll}
\varrho^\pm_{(\iniv_X,\mx{D}_X)(\iniv_Y,\mx{D}_Y)} = \min/\max \sum\limits_{i=1}^{n_X}\sum\limits_{j=1}^{n_Y} \iniv_{X,Y}(i,j) \vect{m}_X(i)\vect{m}_Y(j)-1\\
\mbox{s.t. } \sum\limits_{h=1}^{n_Y} \iniv_{X,Y}(i,h) = \iniv_X(i) \mbox{ and } \sum\limits_{h=1}^{n_Y} \iniv(h,j) = \iniv_Y(j)
\end{array}
\end{equation}
Both LPs depend on different quantities of the APHDs. In the first LP (\ref{eq:lp_in_out}) the spread of the absorption times depending on the exit state of the first random variable and the absorption times depending on the entry state are relevant, whereas the second LP (\ref{eq:lp_input})  considers only absorption times depending on the entry state. 

Computation of an appropriate matrix $\mx{\Phi}_{X,Y}$ or initial vector $\iniv_{X,Y}$ can be achieved by solving linear systems of equations, much more challenging is the construction of appropriate PHDs to allow one to minimize or maximize the correlation for given orders of the PHDs.  Since the first canonical representation has only a single output state, $\mx{a}_X(n_X) =E(X)$, and the second canonical form has a single input state such that $\mx{m}_X(1) = E(X)$, for a sequential composition an APHD in the second canonical form has to be combined with a PHD in the first canonical form and for a parallel composition two APHDs of the first canonical form have to be composed to achieve positive or negative correlation for the bivariate distribution. The following theorems show how to transform a representation of the first canonical form into a representation of the second canonical form. 

\begin{theorem}
\label{th:transform}
For some PHD $(\iniv,\mx{D})$ with the vectors $\mx{m}$, $\exitv$ and $\mx{a}$, a PHD $(\inivp, \mx{D'})$ can be generated by setting $\inivp(i) = \exitv(i)$, $\mu'_{i} =  \mu_{i}$ and 
$$
\mu'_{i,j}=\mu_{j,i} \frac{\statv(j)}{\statv(i)}
$$
then $(\iniv,\mx{D}) \sim (\inivp, \mx{D'})$ and the following relations hold
$$\mx{m'}(i) = \mx{a}(i), \  \exitvp(i) = \iniv(i) \mbox{ and } \mx{a'}(i) = \mx{m}(i).$$
\end{theorem}

If one applies the theorem to an APHD with an upper triangular matrix $\mx{D}$, the resulting matrix $\mx{D'}$ is lower triangular but can be easily transformed into an APHD with an upper triangular matrix by swapping the states which implies that the indices change, i.e., state $i$ in the original APHD corresponds to state $n-i+1$ in the transformed APHD.  The following theorem shows that the theorem transform one canonical form in the other.

\begin{theorem}
\label{th:identity}
If Theorem~\ref{th:transform} is applied to an APHD $(\iniv,\mx{D})$ in canonical form (\ref{eq:first_canonical}), then the resulting APHD $(\inivp,\mx{D'})$ is in canonical form (\ref{eq:second_canonical}) after reversing the order of the states.
\end{theorem}

The theorems imply that if an optimal representation for two PHDs is available, in the sense that it maximizes/minimizes the objective function in (\ref{eq:lp_in_out}) or (\ref{eq:lp_input}), then the representation is also optimal for the other LP after transforming the representation for $X$ using Theorem~\ref{th:transform}. 

We now consider the consecutive generation of APHDs describing exponential distributions. 
To generate an APHD representation of an exponential distribution with dimension $n+1$ from an exponential distribution of dimension $n$, we append  a single phase resulting in a representation $(\inivp,\mx{D'})$. The new APHD is then generated as follows. 
\begin{equation}
\label{eq:stepwise}
\begin{array}{ll}
\inivp = \left((1-p)\iniv, p\right), \ \mx{D'} = \left( \begin{array}{cc}
\mx{D} & \mx{f} \\ \mx{0} & -\mu 
\end{array}\right)\\ \mbox{where } \mx{d}  \ge  \mx{f}, p \in [0, 1)  .
\end{array}
\end{equation}
Let $\mx{d'} = -\mx{D'}\idvt$ and $\inivp_t = \inivp e^{-\mx{D'}t}$. $(\inivp,\mx{D'})$ represents an exponential distribution with rate $\lambda$ if and only if $\frac{\inivp_t\mx{d'}}{\inivp_t\idvts} = \lambda$ for all $t \ge 0$. Again we can assume $\lambda=1$ by scaling the rates in $\mx{D'}$. The following theorem shows that the choice of $\mx{f}$ and $\mu$ underlies severe restrictions.

\begin{theorem}
\label{th:stepwise}
For the representation of an exponential distribution of dimension $n+1$ with rate $1$ from an APHD of an exponential distribution of order $n$ according to (\ref{eq:stepwise}), $\mx{f} = (1-q)\mx{d}$ for $q \in [0,1]$ and $\mu = \frac{1-q+pq}{p}$ are required.
\end{theorem}

Theorem~\ref{th:stepwise} describes an approach to generate the first canonical form as a corner case, where $q=0$ in each step. We will show that this case is optimal if the correlation should be maximized or minimized. Now we consider the vectors $\inivp$, $\exitvp$, $\mx{m'}$ and $\mx{a'}$ which can be generated from the corresponding vectors  $\iniv$, $\exitv$, $\mx{m}$ and $\mx{a}$ of the representation $(\iniv,\mx{D})$ using the following relations.
\begin{equation}
\label{eq:item3}
\begin{array}{llll}
\inivp(i) = \left\{\begin{array}{ll} (1-p)\iniv(i) & \mbox{if } 1 \le i \le n\\ p & \mbox{if } i = n+1\end{array}\right. & 
\exitvp(i) = \left\{\begin{array}{ll}
(1-p)q\exitv(i) & \mbox{if } 1 \le i \le n\\
1-q-pq &\mbox{if } i = n+1
\end{array}\right.\\
\mx{m'}(i) = \left\{\begin{array}{ll}
\mx{m}(i) + \frac{p-pq}{1-q+pq} & \mbox{if } 1 \le i \le n\\
\frac{p}{1-q+pq} & \mbox{if } i = n+1
\end{array}\right. & 
\mx{a'}(i) = \left\{\begin{array}{ll}
\mx{a}(i) & \mbox{if } 1 \le i \le n\\
1 & \mbox{if } i=n+1
\end{array}\right.
\end{array}
\end{equation}
The proof for the equation can be found in the appendix. Theorem~\ref{th:stepwise} shows an approach to expand an APHD by appending one state, it cannot be generate an APHD in the second canonical form. To generate an APHD in the second canonical form we introduce now an approach to prepend a state to an APHD ($\iniv,\mx{D})$ of order $n$ that represents an exponential distribution. The resulting APHD ($\inivp,\mx{D'}$) is of order $n+1$, describes an exponential distribution and is generated as shown in the following equation.
\begin{equation}
\label{eq:stepwise2}
\inivp = \left(p, (1-p)\iniv\right), \ \mx{D'} = \left( \begin{array}{cc} -\mu & (\mu-1)\iniv\\\mx{0} & \mx{D}\end{array}\right) \mbox{ for } \mu \ge 1
\end{equation}
Choosing $\mu\ge \left|\mx{D}(1,1)\right|$ in the $n$th expansion step yields an APHD in the second canonical form. The vectors $\inivp$, $\exitvp$, $\mx{m'}$ and $\mx{a'}$ again can be computed from  $\iniv$, $\exitv$, $\mx{m}$ and $\mx{a}$ using the following relations.
\begin{equation}
\label{eq:item4}
\begin{array}{llll}
\inivp(i) = \left\{\begin{array}{ll} p &  \mbox{if } i = 1\\(1-p)\iniv(i-1) & \mbox{otherwise} \end{array}\right. & 
\exitvp(i) = \left\{\begin{array}{ll}
\frac{p}{\mu} & \mbox{if } i = 1\\
\frac{\mu-p}{\mu}\exitv(i-1) & \mbox{otherwise}
\end{array}\right.\\
\mx{m'}(i) = \left\{\begin{array}{ll}
1 & \mbox{if } i=1\\
\mx{m}(i-1)  & \mbox{otherwise}
\end{array}\right. & 
\mx{a'}(i) = \left\{\begin{array}{ll}
\frac{1}{\mu} & \mbox{if } i = 1 \\
\mx{a}(i-1)+\frac{p(\mu-1)}{\mu(\mu-p)} & \mbox{otherwise}
\end{array}\right.
\end{array}
\end{equation}
The proof for the equation can be found in the appendix.

\section{Positive Correlation}

We consider now APHDs that have been generated according to Theorem~\ref{th:stepwise} and maximize the correlation $\rho^+_{(\iniv,\mx{D})}$. Now let $\rho^{+(n)}$  be the maximal correlation which can be obtained by some APHD  $(\iniv, \mx{D})$ with $n$ phases using the step-wise approach based on Theorem~\ref{th:stepwise}. For a fixed representation the optimal solution results in $\sum_{i=1}^n \iniv(i)\left(\mx{m}(i)\right)^2-1$ where $\mx{m} = \left(-\mx{D}\right)^{-1}\idvt$. Thus, representations have to be found to maximize the  sum. 

We can use the normalized representation with rate $\lambda=1$.  Obviously $\rho^{+(1)} = 0$ because $\mx{D} = (-1)$ and $\iniv=(1)$ in this case. This representation is unique and therefor optimal. Let $\rho^{+(n)}$ be the maximal coefficient of correlation which can be achieved by the step-wise construction of APHDs and let $(\iniv,\mx{D})$ be the corresponding representation.  Now we compute a representation $(\inivp, \mx{D'})$ that achieves $\rho^{+(n+1)}$. According to  (\ref{eq:stepwise}) and Theorem~\ref{th:stepwise}  we have
\begin{equation}
\label{eq:next_m}
\begin{array}{llll}
\inivp = \left((1-p)\iniv, p\right), & \mx{D'} = \left(\begin{array}{cc} \mx{D} & (1-q)\mx{d}\\\mx{0} & -\mu \end{array}\right) & \Rightarrow \\
\mx{M'} = \left(\begin{array}{cc} \mx{M} & \frac{1-q}{\mu}\idvt \\ \mx{0} & \frac{1}{\mu}\end{array}\right), \ &
 \mx{m'} = \left(\begin{array}{c} \mx{m}+\frac{1-q}{\mu}\idvt\\ \frac{1}{\mu}\end{array} \right) 
\end{array}
\end{equation}
where $\mu = \frac{1-q+pq}{p}$.
This results in the following representation of the coefficient of correlation.
\begin{equation}
\label{eq:next_rho}
\begin{array}{lll}
\rho^{(n+1)} & = & \sum\limits_{i=1}^n (1-p)\iniv(i)\left(\mx{m}(i)+\frac{1-q}{\mu}\right)^2 + \frac{p}{\mu^2} - 1\\ 
& = & (1-p) \left(\rho^{+(n)} + 1 + \frac{2(1-q)}{\mu} + \frac{(1-q)^2}{\mu^2}\right) + \frac{p}{\mu^2} - 1 \\
& = & (1-p) \rho^{+(n)} + p\frac{(p(1-p)(1-q)^2}{(1-q+pq)^2}
\end{array}
\end{equation}
The step from the first to the second row results from exploitation of the relations $\rho^{+(n)} = \sum_{i=1}^n \iniv(i)\mx{m}(i)^2-1$ and $\sum_{i=1}^n \iniv(i)\mx{m}(i) = \sum_{i=1}^n\iniv(i) = 1$. The final representation is reached by substituting the representation of $\mu$ from Theorem~\ref{th:stepwise} in the equation. Now consider the second term in the above sum.
\begin{equation}
f(p,q) =  \frac{p(1-p)(1-q)^2}{(1-q+pq)^2} \ \Rightarrow \ \frac{df}{dq} = -\frac{2p^2(1-p)(1-q)}{(1-q+pq)^3} 
\end{equation}
For a fixed $p \in (0,1)$ and $q \in [0,1]$  the first derivative is negative for $q \in [0,1)$ which means that for $q=0$ the maximum is reached for any $p$. Since $q$ does not appear in the first term this also holds for $\rho^{(n+1)}$. Thus, we have
\begin{equation}
\label{eq:opt_rho}
\begin{array}{l}
\rho^{+(n+1)} = \max_{p \in (0,1)} \left((1-p)(\rho^{+(n)}+p)\right)  \\
p = \frac{1-\rho^{+(n)}}{2}=\arg \max_{p \in (0,1)} \left((1-p)(\rho^{+(n)}+p)\right) 
\end{array}
\end{equation}
The resulting APHD is canonical form (\ref{eq:first_canonical}) and $\mu_n = 2/(1-\rho^{+(n)})$. Furthermore,
$\rho^{+(n+1)} = \rho^{+(n)}+ 0.25 \left(1-\rho^{+(n)}\right)^2$ and $\lim_{n \rightarrow \infty} \rho^{+(n)} = 1$. The partial derivatives of $\rho^{(n)}$ with respect to $\mu_i^{-1}$ are given by
\begin{equation}
\label{eq:gradient_max}
\frac{d}{d\mu_i^{-1}} = \iniv(i)\left(\mu_i\mx{m}(i)+2\right)\mx{m}(i) + \sum_{j=1}^{i-1} \iniv(j)\mx{m}(j)\left(2 -\frac{\mx{m}(j)\mu_i}{\mu_i-1}\right)
\end{equation}
If we plug in the values for $\mu_i$ resulting from (\ref{eq:opt_rho}) into (\ref{eq:gradient_max}), the derivatives for $i=2,\ldots,n$ become $0$ such that the necessary conditions for optimality are observed. 

\begin{figure}[htb]
\begin{minipage}[t]{0.48\textwidth}
\includegraphics[width=\textwidth]{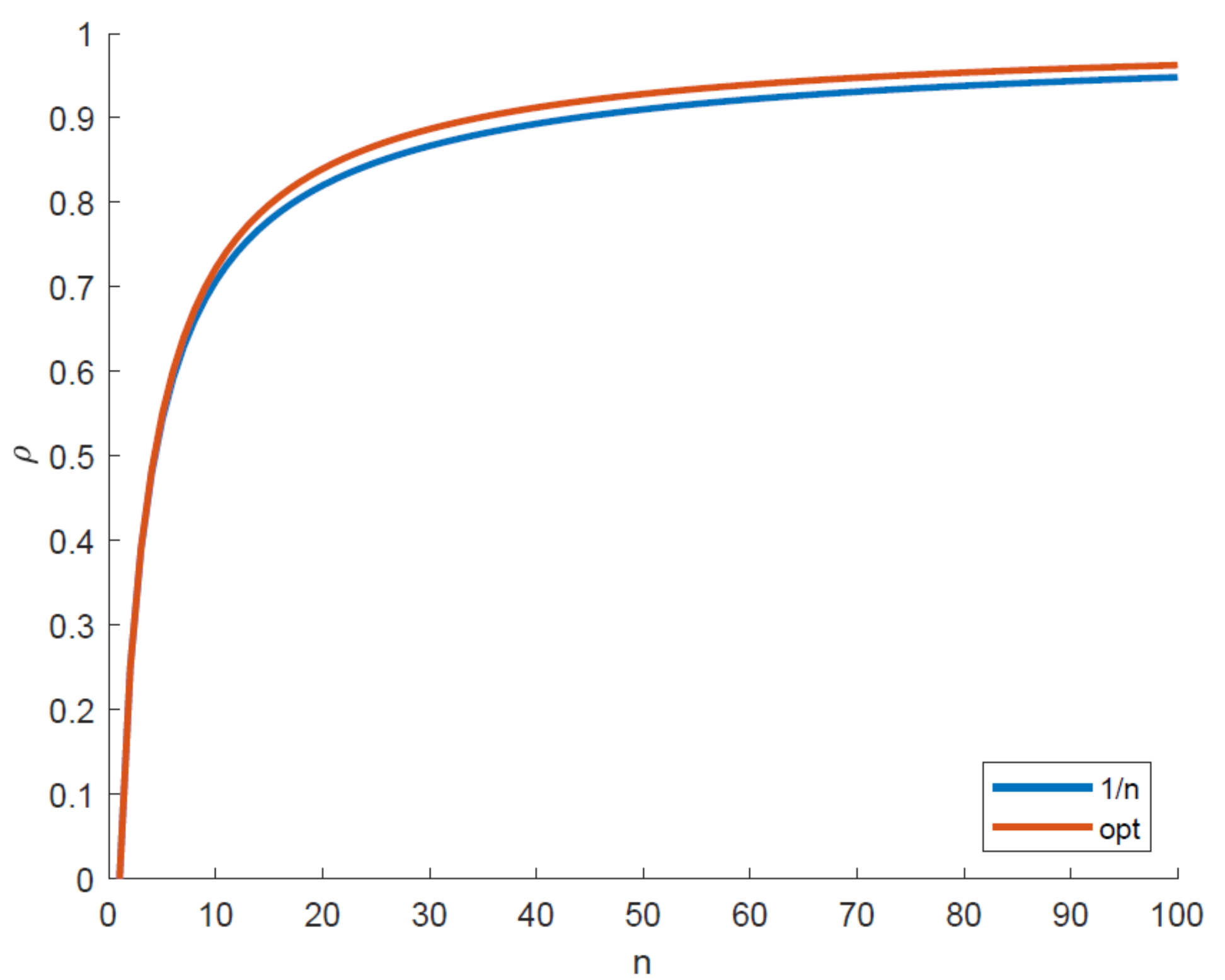}
\end{minipage}
\begin{minipage}[t]{0.48\textwidth}
\includegraphics[width=\textwidth]{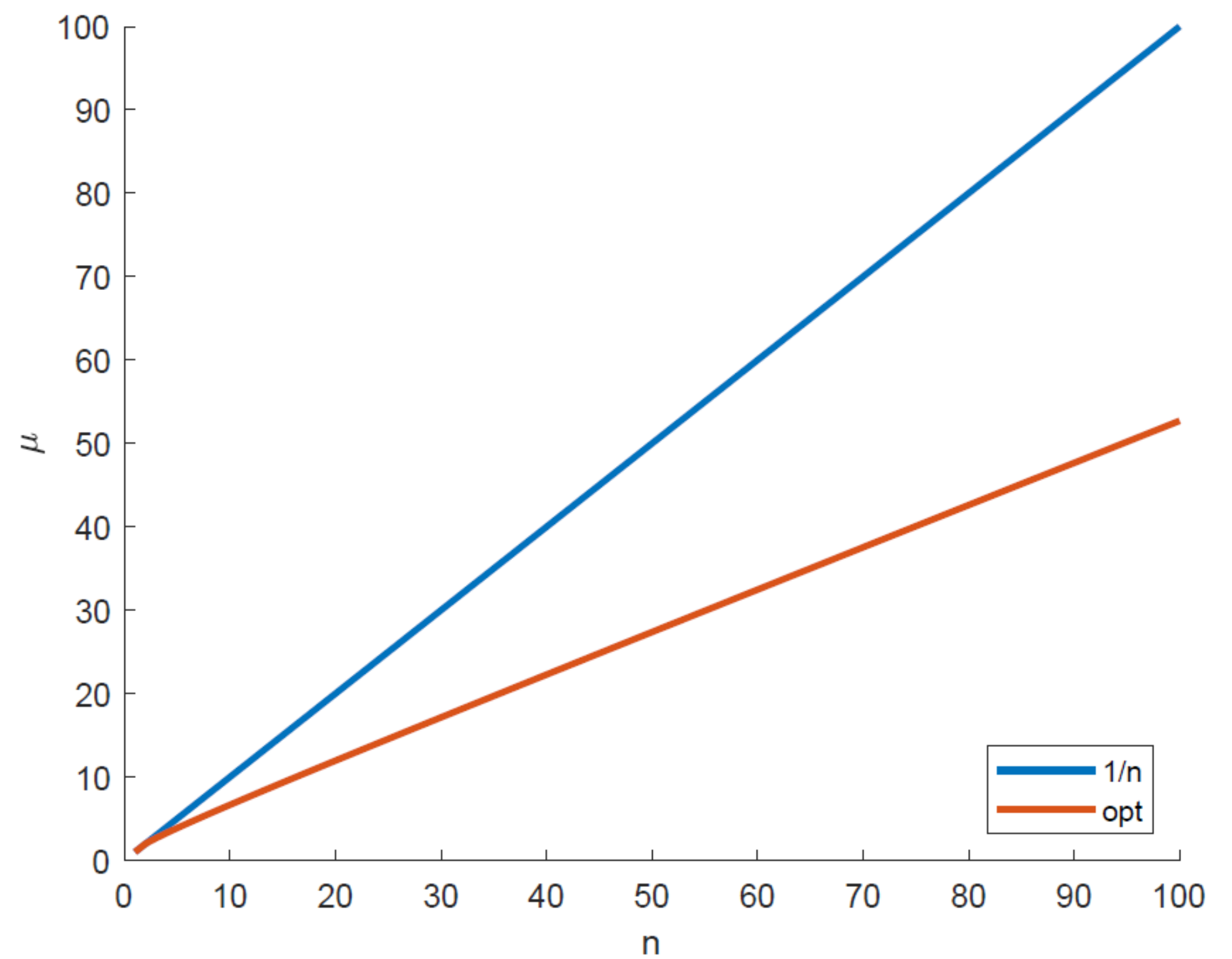}
\end{minipage}
\caption{\label{fig:max_exp} Value of $\rho^{(n)}$ and $\mu_n$ for the representation from (\ref{eq:opt_rho}) and the representations with phase rates $1/i$.}
\end{figure}

We can compare the  generated APHDs  with the representation used in \cite{BlNi10} with $\mu_i = 1/i$. Figure~\ref{fig:max_exp} shows the resulting values for $\rho$ depending on $n$ and the values for $\mu_n$. For both representations $\lim_{n\rightarrow\infty}\rho^{(n)} = 1$ holds. However, the representation from (\ref{eq:opt_rho}) converges slightly faster. This difference is insignificant for smaller values of $\rho$ but becomes significant for $\rho$ near $1$. Thus, for $\rho=0.8$, we need $16$ rather than $18$ phases with the optimized representation. For $\rho=0.9$, $35$ rather than $44$ phases are required, for $\rho=0.95$, $75$ rather than $105$ and for $\rho=0.99$, $393$ rather than $715$ phases. Additionally, rates of the phases remain smaller which might reduce numerical problems or speed up convergence of numerical methods to analyze stochastic systems including the PHDs. 

\section{Negative Correlation}
\label{sec:negative}

We introduce a similar approach to model negative rather than positive correlation between two exponential distributions with normalized APHD representation. In contrast to positive correlation, where for the joint initial vector of two identical exponential APHs $(\iniv_X,\mx{D}_X)$  $\iniv(i,i) = \iniv_X(i)$ holds, it is not obvious which joint initial vector minimizes the correlation. It is known that, if $\iniv(i) = \iniv(n-i+1)$ for all $i=1,\ldots,n$, then $\iniv(i,j)=\iniv(i)$ for $j=n-i+1$ and $0$ otherwise results in the minimal coefficient of correlation because $\mx{m}(i)> \mx{m}(i+1)$. However, the conditions  $\iniv(i) = \iniv(n-i+1)$ puts additional constraints on the class of APHDs and the representation computed for the maximal coefficient of correlation in the previous section does not belong to this class. If the step-wise approach (\ref{eq:stepwise}) with $\mu = \frac{1-q+pq}{p}$ is applied starting with an exponential APHD of order $n$, we obtain the following non-linear program to compute the parameters $p$ and $q$ according to  Theorem~\ref{th:stepwise}. 
\begin{equation}
\begin{array}{ll}
\min_{p,q,\iniv(i,j)} \sum\limits_{i=1}^{n+1} \sum\limits_{j=1}^{n+1} \inivp(i,j)\mx{m'}(i)\mx{m'}(j)\\
\begin{array}{lllll}
s.t. & \mx{m'}(i) = \left\{ \begin{array}{ll}\mx{m}(i) + \frac{p-pq}{1-q+pq}& \mbox{if } i \le n ,\\ \frac{p}{1-q+pq} & \mbox{else,}\end{array}\right. 
& p \in (0,1), \  q \in [0,1] \\
& \sum\limits_{i=1}^{n+1} \inivp(i,j) = \left\{ \begin{array}{ll}(1-p)\iniv(j)  & \mbox{if } i \le n ,\\ p & \mbox{else,}\end{array}\right. &
\sum\limits_{j=1}^{n+1} \inivp(i,j) = \left\{ \begin{array}{ll}(1-p)\iniv(i)  & \mbox{if } i \le n ,\\ p & \mbox{else,}\end{array}\right.\\
\end{array}
\end{array}
\end{equation}
If we fix $p$ and $q$ the problem becomes linear. Solving this problem consecutively for $n=1,2,3,\ldots$ we obtain for the optimal solution $q=0$ and $p = 1/(n+1)$ which results the APHD proposed in \cite{BlNi10}.  In this case $\lim_{n \rightarrow \infty}  \rho^{-(n)} = 1-\pi^2/6$. 

To show that the resulting APHD is not globally optimal, in the sense that we cannot find an APHD representation with $n$ phases that allows us to realize a smaller coefficient of correlation, we now construct an APHD $(\iniv,\mx{D})$ with $3$ phases and the additional restrictions that $\iniv(1)=\iniv(3)$ and that the APHD is in canonical form (\ref{eq:first_canonical}). 
$\iniv(1)=\iniv(3)$ and the canonical representation imply $\left(1-\mu_2^{-1}\right)\left(1-\mu_3^{-1}\right)=\mu_3^{-1}$. Since $\mu_2 < \mu_3$ this results in the condition $\mu_3 > \left(\frac{3}{2}-\sqrt{\frac{5}{4}}\right) \approx 2.618$. The minimal coefficient of correlation for random variables $X,Y$ which are both described by this distribution is reached if $Y$ starts in phase $3$ whenever $X$ starts in phase $1$ and vice versa, additionally both may start in phase $2$. The coefficient of correlation can then be represented in terms of $\mu_3^{-1}$ after some lengthy but simple computations
\begin{equation}
\begin{array}{ll}
\rho_{X,Y} = \frac{1-4\mu_3^{-1}+7\mu_3^{-2}+4\mu_3^{-3}-\mu_3^{-4}}{\left(1-\mu_3^{-1}\right)^2} & \mbox{ and } \\
\frac{d \rho_{X,Y}}{d\mu_3^{-1}} = \frac{-2 +12\mu_3^{-1}-22\mu_3^{-1}+12\mu_3^{-3}+2\mu_3^{-4}-2\mu_3^{-5}}{\left(1-\mu_3^{-1}\right)} .
\end{array}
\end{equation}
The polynomial in the numerator of the derivative has $5$ real valued roots of which only one falls in the interval $(0,0.38197)$ that defines valid values for $\mu_3$. Using this root we obtain $\mu_3=3.09529$, $\mu_2=1.912996$ and $\rho^-_{(\iniv,\mx{D})} = -0.36154$ which is slightly smaller than $\rho^{-(3)} = -0.36111$ which results form $\mu_3=3$ and $\mu_2=2$.  

The presented approach to minimize the coefficient of correlation can hardly be extended beyond $3$ phases because the number of variables in the resulting non-linear functions cannot be optimized in a reliable way. One might apply  algorithms from  nonlinear programming which, however, turn out to be very unreliable and unstable for this class of problems. 

\section{Stochastic Processes wit Exponential Marginal Distribution}
\label{sec:map}

Extensions of PHDs are \textit{Markovian Arrival Processes} (MAPs) \cite{Neut79,Luca91}. A MAP is characterized by two $n \times n$ matrices $(\mx{D}_0,\mx{D}_1)$ where $\mx{D}_0$ is a sub-generator, $\mx{D}_1$ is non-negative and $\mx{D}_0+\mx{D}_1$ is an irreducible generator with stationary vector $\statv$. Let $\mx{P} =-\mx{D}_0^{-1}\mx{D}_1$ which is an irreducible stochastic matrix and $\iniv\mx{P}=\iniv$ is the unique left eigenvector for eigenvalue $1$. Then $(\iniv,\mx{D}_0)$ is the PHD describing the marginal distribution of the MAP \cite{BKF14}. One can also expand a PHD to a MAP by choosing $\mx{D}_1$ such that $\iniv$ becomes the left eigenvector of the resulting matrix $\mx{P}$. This approach is applied in so called two-phase moment fitting for MAPs \cite{BKF14,HoTB05}. The coefficient of autocorrelation of a MAP with an embedded APHD that describes a normalized exponential distribution is given by
\begin{equation}
\rho_{(\mx{D}_0,\mx{D}_1)} = \iniv\mx{D}_0^{-1}\mx{P}\mx{D}_0^{-1}\idvt - 1. 
\end{equation}
 In this section we consider a MAP expansion of an APHD representation of an exponential distribution that minimizes or maximizes the coefficient of autocorrelation. 

Assume that $(\iniv, \mx{D})$ is an APHD representation of the exponential distribution with rate $1$, then this representation can be expanded to a MAP by defining probabilities $\intv(i,j)$. To maximize/minimize the coefficient of correlation, the following problem has to be solved.  
\begin{equation}
\label{eq:map_lp}
\begin{array}{ll}
\min/\max \sum\limits_{i=1}^n\sum\limits_{j=1}^n \intv(i,j) \mx{a}(i)\mx{m}(j) - 1\\
\mbox{s.t. }  \sum\limits_{i=1}^n \intv(i,j) = \iniv(j) \mbox{ and } \sum\limits_{j=1}^n \nu(i,j) = \exitv(i)
\end{array}
\end{equation}
Matrix $\mx{D}_1$ is then given by $\mx{D}_1(i,j) = \left(-\sum_{k=1}^n\mx{D}_0(i,k)\right)\left(\sum_{l=1}^n \intv(i,l)\right)^{-1}\intv(i,j)$ for $\intv(i,j) > 0$ (i.e., $\exitv(i) > 0$)  and $0$ otherwise. (\ref{eq:map_lp}) is a linear program, if the vector $\mx{a}$ and $\mx{m}$ are known. Both vectors are determined by the representation $(\iniv,\mx{D})$.

\begin{theorem}
\label{th:map}
A MAP with an acyclic matrix $\mx{D}_0$ and an exponential marginal distribution has to be at least of order $3$ to model non-zero autocorrelation.
\end{theorem}

The canonical representations for exponential APHDs cannot be used as matrices $\mx{D}_0$ for a MAP to model correlation because the representations either have a single input or a single output state which implies that subsequent events are uncorrelated. However, each canonical representation of order $n$ describes $n$ paths, ($\mu_n$),  ($\mu_{n-1},\mu_n$), $\ldots$, e.g., ($\mu_1,\ldots,\mu_n$) are the paths of the first canonical representation and $\pathv_1=\iniv(n),\ldots,\pathv_n=\iniv(1)$ are the probabilities to choose these paths. Thus, for an APHD $(\iniv,\mx{D})$  in the first  or second canonical form, an expanded APHD $(\inivp, \mx{D'})$ with $n \frac{n+1}{2}$ states can be can be generated that contains all the paths in isolation, i.e., for each path a unique entry and a unique exit state exist. 
\begin{equation}
\label{eq:map_from_ph}
\inivp = \left(\pathv_1,\pathv_2,0,\pathv_3,\ldots,\pathv_n,0,\ldots,0\right),
\mx{D'} = \left(\begin{array}{cccccc}
-\mu_n & \\
 & -\mu_{n-1} & \mu_{n-1}\\
& & -\mu_n \\ & & & \ddots\\ & & & & -\mu_n
\end{array}\right)
\end{equation}
Let $e_i$ and $f_i$ be the entry and exit state ,respectively of the $i$th path. $e_i$ are the only entry states and $f_i$ are the  only exit states of the distribution ($i=1,\ldots,n$). Furthermore, $\mx{m'}(e_i) = \mx{a'}(f_i)=\mx{m}(n-i+1)$ where $\mx{m}$ is the moment vector of APHD in the first canonical form with rates $\mu_1,\ldots,\mu_n$. 

\begin{theorem}
\label{th:map_moments}
If $(\iniv,\mx{D})$ is an APHD representation of an exponential distribution with minimal/maximal coefficient of correlation $\rho^-_{(\iniv,\mx{D})}$ and $\rho^+_{(\iniv,\mx{D})}$, then the minimal/maximal coefficient of autocorrelation of the a MAP resulting from $(\inivp, \mx{D'})$ computed as in (\ref{eq:map_from_ph}) equals $\rho^-_{(\iniv,\mx{D})}$ and the maximal coefficient of autocorrelation equals $\rho^+_{(\iniv,\mx{D})}$. 
\end{theorem}

 Thus, MAPs with exponential marginal distribution and an arbitrary coefficient of autocorrelation in the possible range can be generated using an extension of the idea proposed by \cite{BlNi10} (see also Section~\ref{sec:related}) can be generated. Unfortunately,  the number of phases grows quadratically compared to the number of phases required for bivariate exponential distributions. 
 
\section{Beyond Exponential Distributions}
\label{sec:general}

Since exponential distributions are the base for PHDs, it is possible to present single phases of an arbitrary PHD by APHDs to increase or decrease the coefficient of correlation that can be reached by correlated PHDs or MAPs. The approach is introduced here for hyperexponential or Erlang distributions but may be applied for other types of PHDs as well.

\subsection{Hyperexponential Distributions}

Hyperexponential distributions are APHDs where matrix $\mx{D}$ is a diagonal matrix. It is known that hypexponential distributions with only two phases can be applied to describe distributions with arbitrary, but finite, first moment and coefficient of variation larger than $1$. The moments of the hyperexponential distribution with $n$ phases are given by 
\begin{equation}
E(T^i) = i!\sum_{j=1}^n \frac{\iniv(j)}{\mu_j^i}
\end{equation}
where $E(T^i)$ is the $i$th moment of the distribution. The $2n-1$ parameters of the distribution can then be selected to match or approximate some quantities like higher order moments or values of the CDF from some real process \cite{FeWh98,RiDS04}. For a hyperexponential distribution $\mx{m}(i) = \mx{a}(i)$ and $\iniv(i) = \exitv(i)$. Without loss of generality we can assume that $E(T^1)=1$ and $\mu_1 < \mu_2 < \ldots < \mu_n$. 

The maximal/minimal coefficient of correlation of two combined hyperexponential distributions  is again given by the solution of the LP in (\ref{eq:lp_input}) which can be achieved by consecutively assigning probabilities as it has been done done for APHDs in canonical form. Thus, for a hyperexponential distribution $(\iniv,\mx{D})$ of order $n$, $\rho^+_{(\iniv,\mx{D})}$ is given by
\begin{equation}
\label{eq:hyperexp_cc}
\rho^+_{(\iniv,\mx{D})} = 
\frac{\sum\limits_{i=1}^n\frac{\iniv(i)}{\mu_i^2}-\left(\sum\limits_{i=1}^n\frac{\iniv(i)}{\mu_i}\right)^2}
{2\sum\limits_{i=1}^n\frac{\iniv(i)}{\mu_i^2}-\left(\sum\limits_{i=1}^n\frac{\iniv(i)}{\mu_i}\right)^2} = 
\frac{\sigma^2-\frac{1}{2}E(T^2)}{\sigma^2} = 
\frac{1}{2}\cdot\frac{\sigma^2-E(T)^2}{\sigma^2} < \frac{1}{2}
\end{equation} 
where $E(T),E(T^2)$ are the first two moments and $\sigma^2$ is the variance of the hyperexponential distribution.  This result has been derived for hyperexponential distributions with $2$ phases in \cite{HeML03} but also holds for more than two phases. 
The maximal coefficient of correlation only depends on the first two moments and not on the parameters of the hyperexponential distribution and it is always smaller than $0.5$. To increase the coefficient of correlation each exponential phase of the hyperexponential distribution can be substituted by one of the APHD representations of the exponential distribution proposed above. Assume that phase $i$ with rate $\mu_i$ is substituted by an APHD in the first canonical form with $n_i$ phases. Let $\mx{m}_i$ be the moment vector of the APHD representation of the expansion of phase $i$ of the hyperexponential distribution and $\iniv_i$ the corresponding initial vector, both vectors are of length $n_i$. To obtain maximal correlation of two hyperexponential distributions with an identical representation, both distributions are started jointly in phase $i$ of the hyperexponential distribution and phase $j$ ($\in \{1,\ldots,n_i\}$) of the APHD representation of the $i$th phase with probability $\iniv(i)\iniv_i(j)$ where $\iniv_i(j)$ is the probability of starting the APHD representing the $i$th exponential phase of the hyperexponential distribution. The coefficient of correlation is then given by
\begin{equation}
\label{eq:cc_expanded}
\frac{\sum\limits_{i=1}^n\iniv(i)\sum\limits_{j=1}^{n_i}\iniv_i(j)(\mx{m}_i(j))^2-E(T)^2}{\sigma^2}
\end{equation}

\begin{theorem}
\label{th:hyperexp}
If for a hyperexponential distribution $(\iniv,\mx{D})$ with $n$ phases each phase $i=1,\ldots,n$ is substituted by an APHD in canonical form according to (\ref{eq:opt_rho}) resulting in APHD $(\inivp,\mx{D'})$, then 
\begin{equation}
\label{eq:hyperexp}
\rho^+_{(\inivp,\mx{D'})} = \frac{\sum\limits_{i=1}^n \frac{\iniv(i)\rho^{+(n_i)}}{\mu_i^2}+\frac{1}{2}E(T^2)-E(T)^2}{\sigma^2} .
\end{equation}
This implies $\lim_{n_1,\ldots,n_n \rightarrow \infty}\rho^+_{(\inivp,\mx{D'})} = 1$. 
\end{theorem}

The proposed approach to combine two identical hyperexponential distributions according to joint initial probabilities can be extended as described for exponential distributions above. Theorem~\ref{th:transform} can be applied to transform phases expanded according to  the first canonical form into a hyperexponential distribution where phases are expanded according to the second canonical form.  Then the two distributions can be combined sequentially as shown in (\ref{eq:expansion_out_in}). Hyperexponential distributions with different number of phases can be combined in exactly the same way, the joint initial vector or the matrix $\mx{\Psi}$ have then computed from the LP problem (\ref{eq:lp_input}) or (\ref{eq:map_lp}). For $n_i \rightarrow \infty$ the maximal coefficient of correlation converges toward $1$. 

To generate a phase type representation of a hyperexponential distribution of order $n$ that reaches a given coefficient of correlation $\rho > 0$ and a minimal number of phases, the exponential phases are expanded by adding in each step a phase starting with the original hyperexponential distribution with $n$ phases. Now assume that a representation $(\iniv_X,\mx{D}_X)$ has been reached where phase $i$ has been expanded into an APHD of order $n_i$, then (\ref{eq:hyperexp}) holds. 
If $\rho_{(\iniv,\mx{D})}^+ \ge \rho$ we are done, otherwise we choose one phase of the hyperexponential distribution and add another phase in the corresponding APHD (i.e., $n_i \rightarrow n_i+1$) resulting in a new PHD $(\iniv^{(i)},\mx{D}^{(i)})$. The coefficient of correlation then becomes
\begin{equation}
\rho_{(\iniv^{(i)},\mx{D}^{(i)})}^+  = \frac{\sum\limits_{j=1,j\neq i}^n \iniv(j)\frac{\rho^{+(n_j)}}{\mu_j^2}+\iniv(i)\frac{\rho^{+(n_i+1)}}{\mu_i^2}+\frac{1}{2}E(T^2)-E(T)^2}{\sigma^2} =
\rho_{(\iniv,\mx{D})}^+ + \frac{\iniv(i)\frac{\rho^{+(n_i+1)}-\rho^{+(n_i)}}{\mu_i^2}}{\sigma^2} .
\end{equation}
To keep the number of states in the resulting PHD small, phase $i$ of the hyperexponential distribution that maximizes $\iniv(i)\frac{\rho^{+(n_i+1)}-\rho^{+(n_i)}}{\mu_i^2}$ is chosen in the next expansion step. 

The same approach can also be applied for negative correlation using the results from Section~\ref{sec:negative} for the expansion of the phases. Furthermore, the approach can be applied to build MAPs with hyperexponentially distributed marginal distribution by expanding the phases of the hyperexponentially distribution according to the results in Section~\ref{sec:map}. 

\subsection{Erlang Distributions}

Results are available to generate random variates of correlated Erlang distributions \cite{BaLa09} but little is known how correlated Erlang distributions can be described by PHDs. We present a first approach here which, unfortunately, requires a huge number of phases to obtain a small or a large coefficient of correlation for an Erlang distribution of a higher order. An Erlang $k$ distribution consists of $k$ phases each with rate $\lambda$ that are all visited consecutively such that for a random variable $X$ with Erlang $k$ distribution $E(X) = \frac{k}{\lambda}$, $E(X^2)=\frac{k(k+1)}{\lambda^2}$ and $\sigma^2_X = \frac{k}{\lambda^2}$. We consider again the normalized variant with $E(X) = 1$ and $\lambda = k$. 

To model correlation between Erlang distributions described by PHDs, each exponential phase is substituted by a set of paths resulting from the canonical representations. A canonical representation of order $n$ describes $n$ paths. If each phase of the Erlang distribution is substituted by an APHD with $n$ phases, $n^k$ possible paths through the Erlang distribution exist. However, the ordering in which the exponential phases on a path are passed is irrelevant.  Thus elementary combinatorics shows that we only need to consider ${n + k - 1\choose k}$ different paths. These paths contain between $k$ and $kn$ phases. For $n=2$ and phase type representations of the first canonical form, we have the paths $(\mu_1,\mu_2)$ which is chosen with probability $\iniv(1)$ and $(\mu_2)$ which is chosen with probability $\iniv(2)$. The corresponding phase representation $(\inivp. \mx{D'})$  of an Erlang $2$ distribution looks as follows.
\begin{equation}
\label{eq:erlang2_all}
\inivp = \left(\begin{array}{c} \iniv(1)^2\\ 0\\ 0 \\ 0\\2\iniv(1)\iniv(2)\\ 0\\ 0\\ \iniv(2)^2\\ 0 \end{array}\right)^T, \
\mx{D'} = \left( \begin{array}{ccccccccc}
-\mu_1 & \mu_1 & 0 & 0 & 0 & 0 & 0 & 0 & 0\\
0 & -\mu_2 & \mu_2 & 0 & 0 & 0 & 0 & 0 & 0 \\
0 & 0 & -\mu_1 & \mu_1 & 0 &0 &0 &0 &0 \\
0 & 0 & 0 & -\mu_2 & 0 & 0 &0 & 0 & 0\\
0 & 0 & 0 & 0 & -\mu_1 & \mu_1 & 0 &0 &0 \\
0 & 0 & 0 & 0 & 0 & -\mu_2& \mu_2 &0 & 0 \\
0 & 0 & 0 & 0 & 0 & 0 & -\mu_2 & 0 & 0 \\
0 & 0 & 0 & 0 & 0 & 0 & 0 & -\mu_2 & \mu_2\\
0 & 0 & 0 & 0 & 0 & 0 & 0 & 0 & -\mu_2
\end{array}\right)
\end{equation}
This representation of the Erlang distribution can be used to define matrix $\mx{D}_0$ of a MAP with Erlang distributed marginal distribution. If Erlang distributions are combined in parallel, exit states are not relevant. Thus common postfixes (in terms of the exponential phases)  of paths  can be combined. Thus, the following representation of the Erlang 2 distribution allows one to describe the same correlation coefficients than (\ref{eq:erlang2_all}) in a parallel composition but has only $4$ rather than $9$ phases.
\begin{equation}
\label{eq:erlang_in}
\inivp = \left(\begin{array}{c} \iniv(1)^2\\ 2\iniv(1)\iniv(2)\\ \iniv(2)^2\\ 0 \end{array}\right)^T, \
\mx{D'} = \left( \begin{array}{ccccccccc}
-\mu_1 & \mu_1 & 0  & 0 \\
0 & -\mu_1 & \mu_1  & 0 \\
0 & 0 &  -\mu_2 & \mu_2  \\
0 & 0 &  0 &  -\mu_2
\end{array}\right)
\end{equation}
The reduction can be applied for arbitrary $k$ and $n$. The representation has a single output state and ${n + k - 1\choose k}$ input states. For $k > 2$ there are $k-2$ additional states that are not input and not output states. If we use the lexicographical ordering of the paths, then the states have decreasing conditional sojourn times, because $\mu_i < \mu_{i+1}$. The corresponding representation with a single input and  ${n + k - 1\choose k}$ output states is shown in the following equation.
\begin{equation}
\label{eq:erlang_out}
\inivp = \left(\begin{array}{c} 1 \\ 0 \\ 0 \\ 0\end{array}\right)^T, \
\mx{D'} = \left( \begin{array}{ccccccccc}
-\mu_2 & \mu_2 & 0  & 0 \\
0 & -\mu_2 & (1-\iniv(1))^2\mu_2  & 0 \\
0 & 0 &  -\mu_1 & \frac{2\iniv(1)\iniv(2)}{2\iniv(1)\iniv(2)+\iniv(2)^2}\mu_1  \\
0 & 0 &  0 &  -\mu_1
\end{array}\right)
\end{equation}
This representation has a single input and $4$ output states. To combine two Erlang distributions sequentially, the first has a representation as in (\ref{eq:erlang_out}) and the second a representation as in (\ref{eq:erlang_in}). 

\section{Examples}

We consider two simple queueing models to show the effect of correlation on the behavior of systems.

\subsection{An M/M/1 System with Correlation between Inter-Arrival and Service Time}

Single queues with correlated arrivals and  correlated service times can be modeled as MAP/PH/1, PH/MAP/1 or MAP/MAP/1 queues. More recently it has been shown that also correlations between inter-arrival and service times can be modeled by PHDs and analyzed with matrix geometric methods \cite{BK17,He12,Houdt12,LHB06}. To model correlation of arrival and service times, a PH/PH/1 queue with $K$ customer classes is considered, where $K$ is the number of exit states of the PHD describing arrivals. For each class a different initial vector for the PHD describing the service times is defined. To model an M/M/1 queue with correlated arrival and service times and correlation coefficient $\rho>0$ we first have to find the minimal $n$ such that $\rho \le \rho^{+(n)}$. Then the corresponding APHD of the first canonical form is generated and used to represent the service time, the distribution is also transformed to the second canonical form to represent the arrivals. Both distributions are multiplied with the corresponding rates to have the correct arrival and service rates. A customer that arrives from phase $k$ of the arrival time distribution is assigned to class $k$, the initial vector of class $k$ is chosen as $\rho/\rho^{+(n)}\mx{e}_i+(1-\rho/\rho^{+(n)})\iniv$ where $\iniv$ is the initial vector of the computed APHD in the first canonical form. Negative correlations can be represented similarly. The resulting queue can be analyzed with the algorithms available in \cite{BMSPH12}. 

\begin{figure}[htb]
\begin{minipage}[t]{0.48\textwidth}
\includegraphics[width=\textwidth]{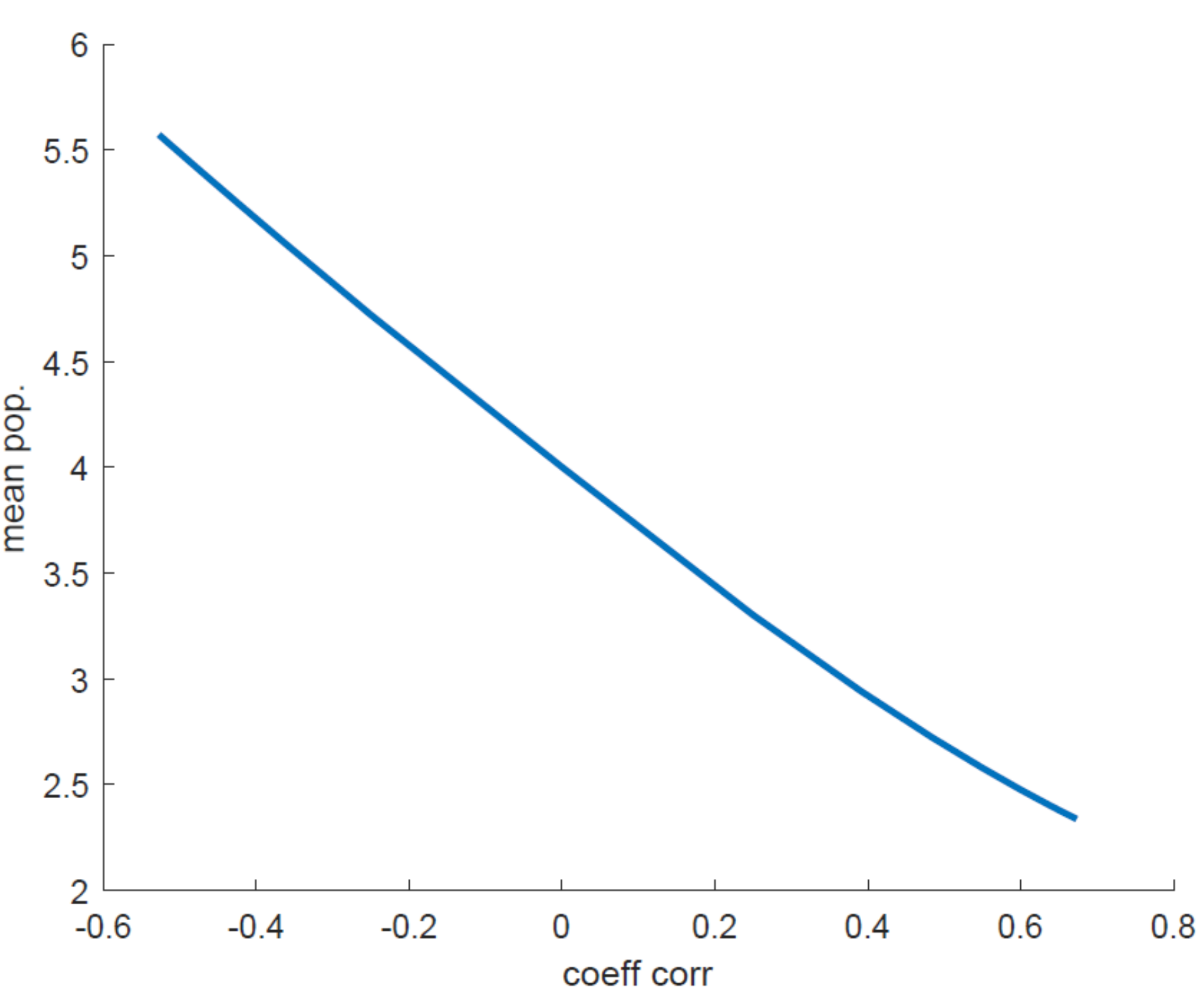}
\end{minipage}
\begin{minipage}[t]{0.48\textwidth}
\includegraphics[width=\textwidth]{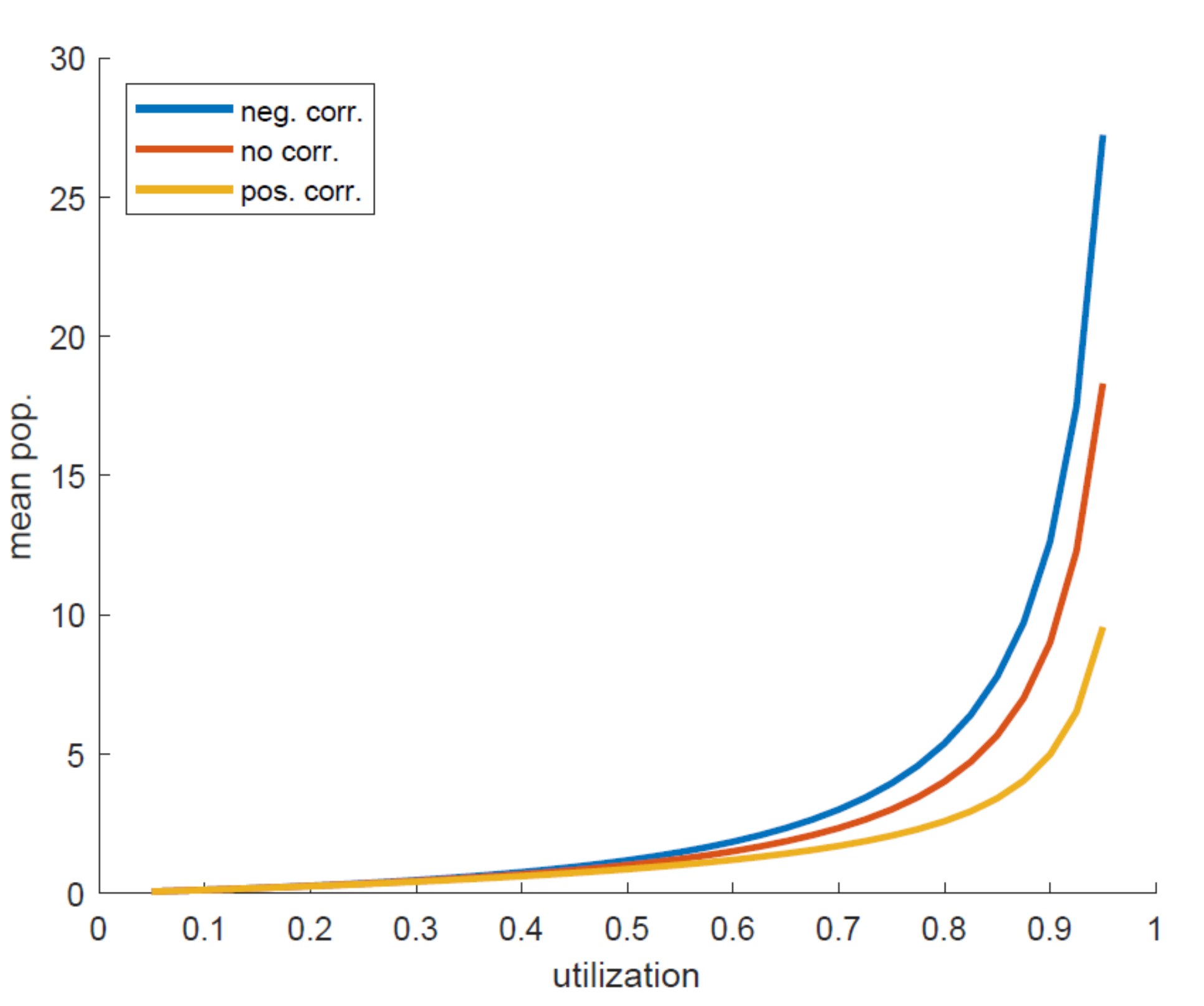}
\end{minipage}
\caption{\label{fig:mm1_pop} Mean population of an M/M/1 queue with correlation between arrival and service times. Left side mean queue length depending on the coefficient of correlation for utilization $0.8$. Right side mean population for varying utilization and negative correlation (coefficient of correlation $-0.4636$), no correlation and positive correlation (coefficient of correlation $0.5502$)}
\end{figure} 

\begin{figure}[htb]
\begin{minipage}[t]{0.48\textwidth}
\includegraphics[width=\textwidth]{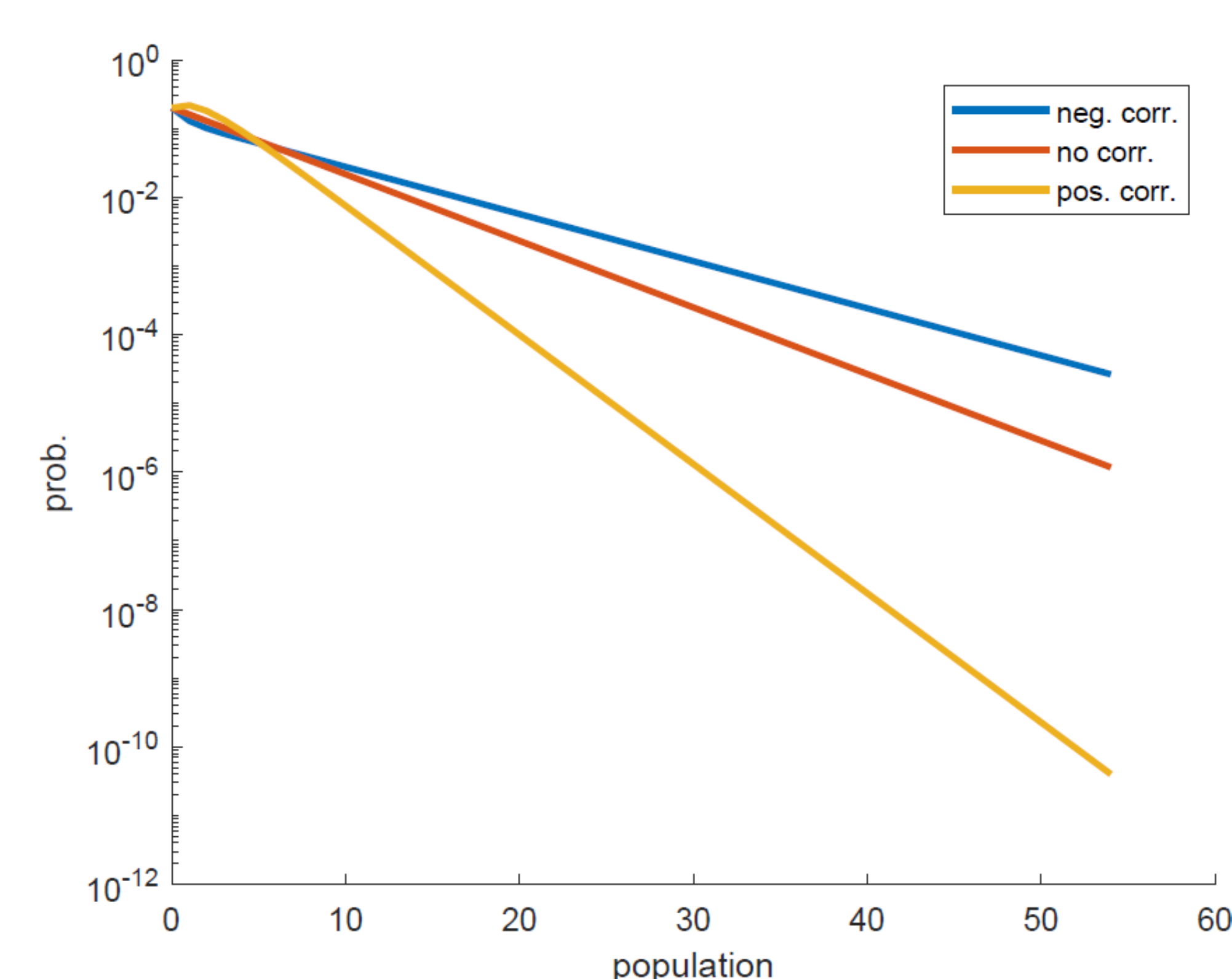}
\end{minipage}
\begin{minipage}[t]{0.48\textwidth}
\includegraphics[width=\textwidth]{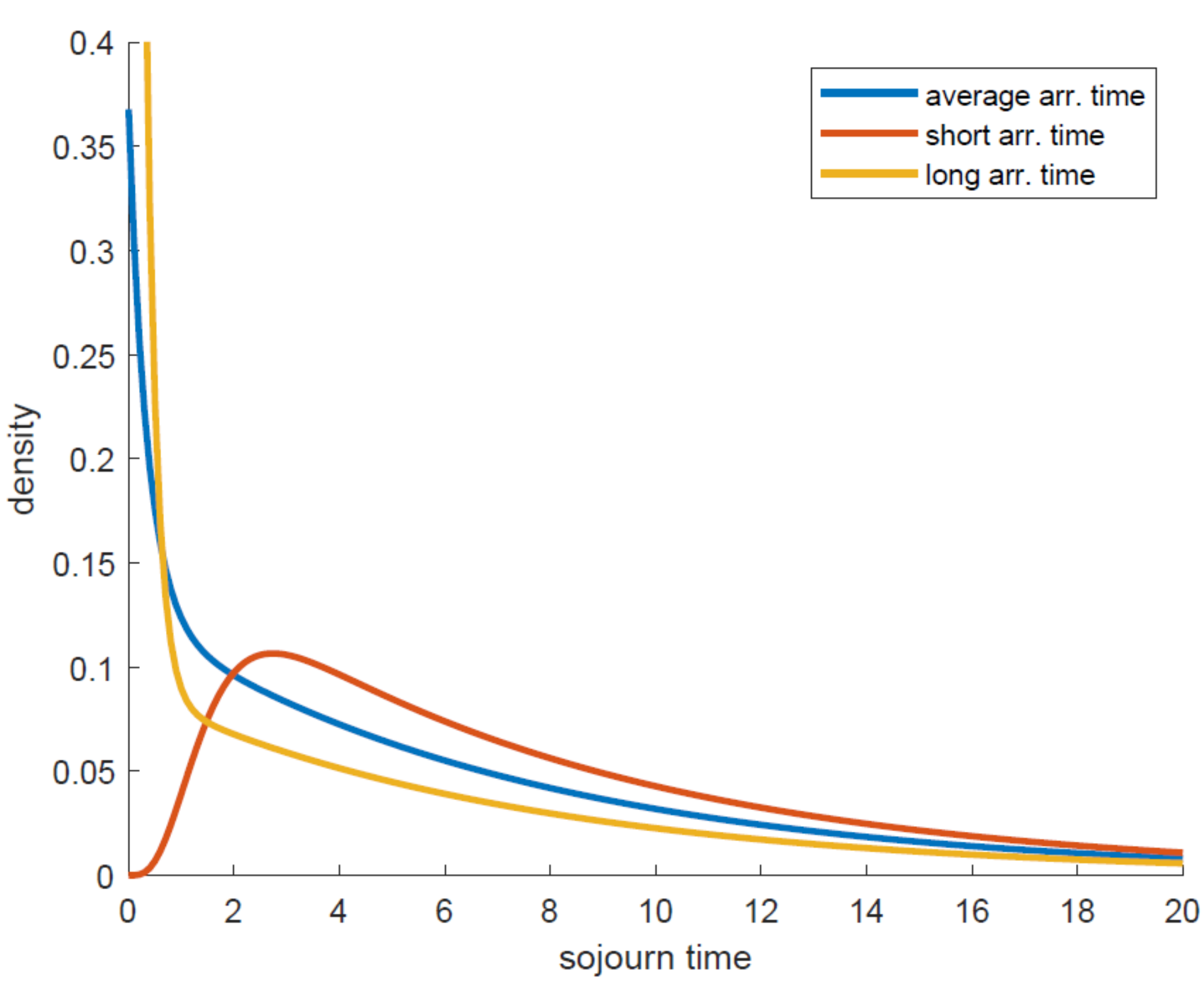}
\end{minipage}
\caption{\label{fig:mm1_detail} M/M/1 queue with utilization $0.8$ and mean service time $1$. Left side queue length distribution for negative correlation, positive correlation or no correlation. Right side density of the sojourn time for the queue with negative correlation   (coefficient of correlation $-0.4636$). Density for average customers, customers departing from the first phase of arrival time distribution and arrivals from the last phase of the arrival time distribution.}
\end{figure}

We present some results for an M/M/1 queue where inter arrival and service times are correlated. Figure~\ref{fig:mm1_pop} shows the mean population for different versions of the queue. It can be seen that a negative correlation between both values results in a significant larger population , in particular for a larger utilization, and that the population depends for a given utilization almost linearly on the coefficient of correlation. Some more detailed results can be found in Figure~\ref{fig:mm1_detail}. It can be noticed that correlation has a significant influence on the tail of the distribution of queue length and that the sojourn time depends heavily on the inter arrival time, if inter-arrival and service time are negatively correlated. 

\begin{table}
\begin{center}
\begin{tabular}{|c|c|c|c|c|c|}\hline
& \multicolumn{5}{|c|}{$U$}\\\hline
$\rho$   &  0.2 & 0.5 & 0.8 & 0.9 & 0.95\\\hline
-0.5498 & 828.9 & 1733.5 & 4372.1 & 8263.5 & 15609\\
-0.4636 & 0.95   & 1.82    & 5.48     & 10.46  & 20.51 \\
-0.2500 & 0.00   & 0.01    & 0.03     & 0.03    & 0.07\\
 0.2500 & 0.00   & 0.00    & 0.01     & 0.02    & 0.03\\
 0.6008 & 0.71   & 1.31    & 2.24     & 4.07    & 6.90\\
 0.7222 & 622.2 & 994.2  & 1608.9  & 2363.9 & 3862.7\\\hline 
\end{tabular}
\caption{\label{tab1} CPU times in seconds to solve the M/M/1 queue with varying coefficient of correlation $\rho$ and utilization $U$.}
\end{center}
\end{table}

The solution algorithm required to solve the M/M/1 queue is based on the solver for the solution of SM[K]/PH[K]/1/FCFS queues from \cite{He12}, a matlab implementation of this algorithm is available in \cite{BMSPH12}. Unfortunately, the block size of the matrix blocks solved in the matrix geometric solution is in $O(n^3)$ where $n$ is the dimension of the APHD representation of the exponential distributions. Thus, for small and large coefficient of correlation, the effort grows quickly since the number of phases grows quickly as shown in Figure~\ref{fig:max_exp}. Table~\ref{tab1} includes the cpu times required to solve the queue on a standard PC with Intel(R) Core(TM) i5-9400 CPU @ 2.90GHz with 6 cores  and 8 GB of main memory. The wall clock time is shorter because matlab already uses some parallel steps. It can be noticed that the solution time mainly depends on the number of phases. In the example we used $2$, $5$ and $10$ phases. Since the block size for the matrix geometric solution grows with $O(n^3)$, it is obvious that it becomes very costly to increase or decrease the coefficient further. Additionally, solution time grows with the utilization of the queue because the solver requires more iterations to converge and solution takes longer for negative than for positive correlation with the same number of phases because the rates are larger in the negative case.

\subsection{A Queue with Correlated Task Processing Times}

We now consider a queue with a Poisson arrival process where jobs consist of a pair of tasks. Tasks are processed on a single processor, processing times are exponentially distributed but might be correlated. Without correlation the model is a simple $M/E_2/1$ queue.  To model correlation the first phase of the Erlang $2$ distribution is expanded to an APHD in the second canonical form and the second phase into an APHD in the first canonical form. Both APHDs are combined sequentially. The number of phases depends on the required coefficient of correlation. The resulting model is an $M/GI/1$ or better $M/PH/1$ queues with the same mean service time than the $M/E_2/1$ queue modeling the uncorrelated case but with a different second moments of the service time. For a stationary analysis we only have to determine the second moments of the absorption time of the sequentially combined APHDs (cf. (\ref{eq:expansion_out_in})) to apply the PK-formula for $M/GI/1$ systems. 

\begin{figure}[htb]
\begin{minipage}[t]{0.48\textwidth}
\includegraphics[width=\textwidth]{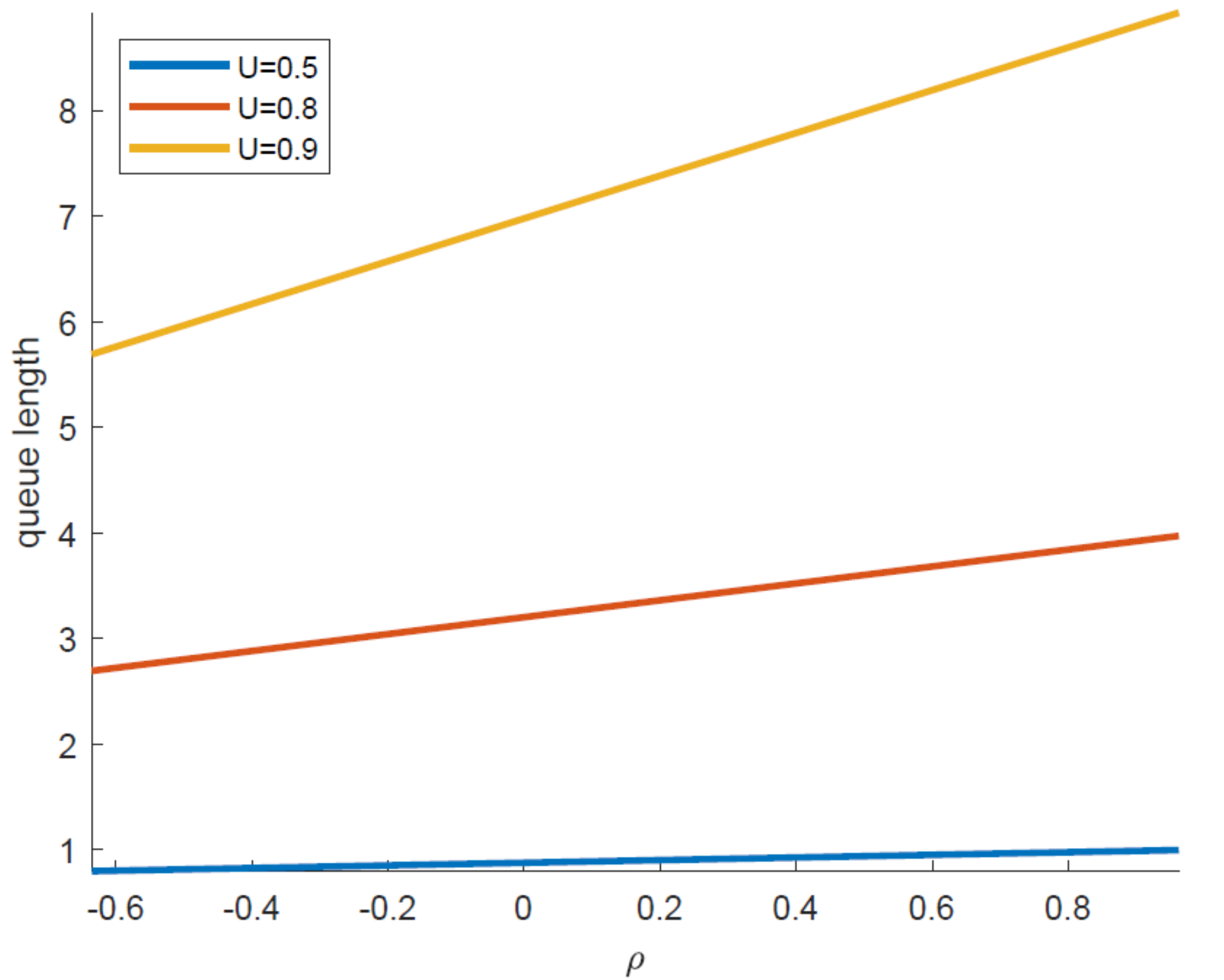}
\end{minipage}
\begin{minipage}[t]{0.48\textwidth}
\includegraphics[width=\textwidth]{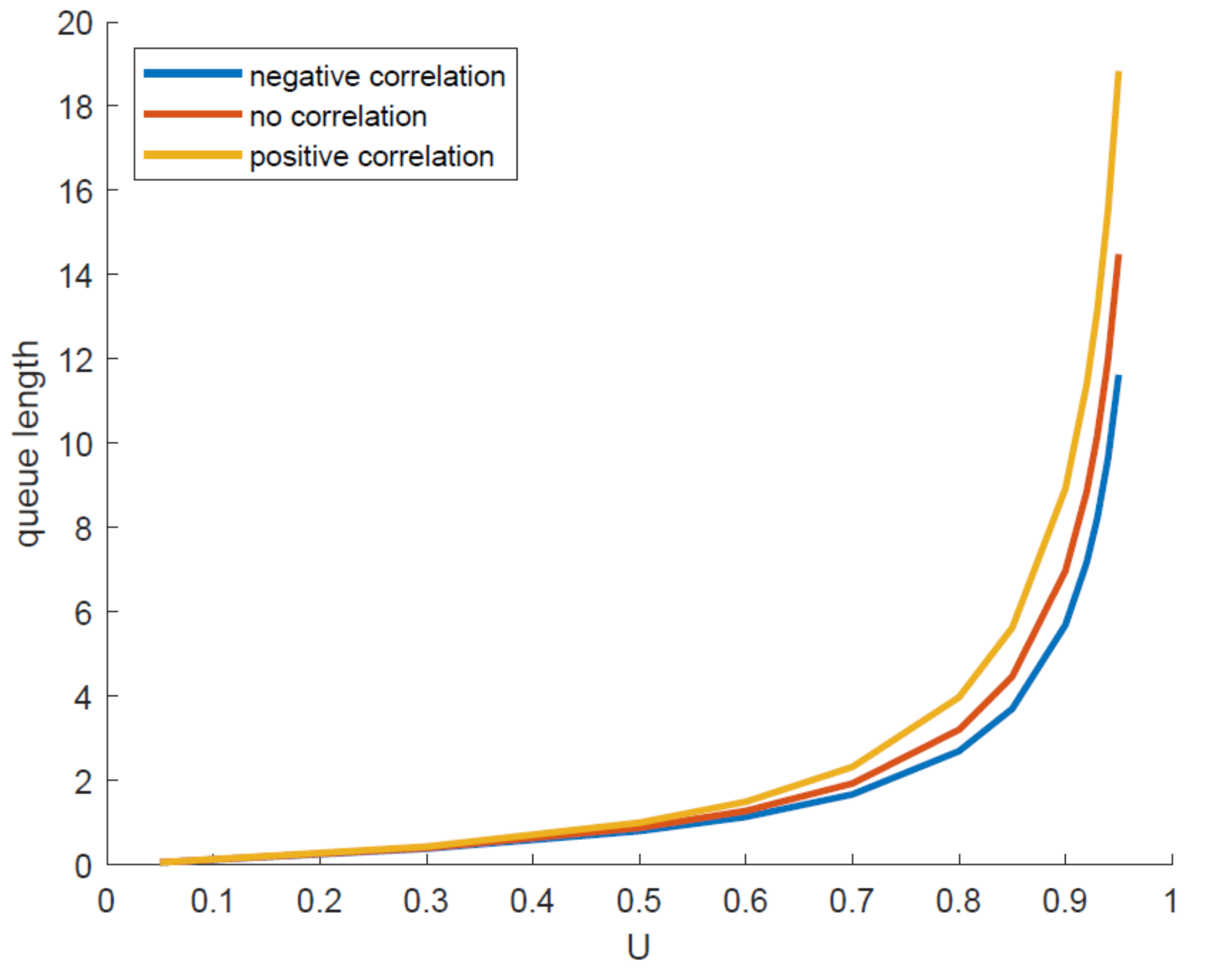}
\end{minipage}
\caption{\label{fig:mg1_pop} Queue with two correlated tasks. Left side mean population in the queue for different coefficients of correlation $\rho$ and varying utilization $U$. Right side mean population in the queue for varying utilization, low correlation ($\rho=-0.635$), no correlation ($\rho=0$) and high correlation ($\rho = 0.962$). }
\end{figure} 

Figure~\ref{fig:mg1_pop} shows some results for the queue. It can be noticed that again the mean population depends almost linearly on the coefficient of correlation and the effect is larger for a higher utilization. In this configuration a positive correlation results in a larger coefficient of correlation of the service time and thus a larger population in the queue. Furthermore, the right figure shows that the effect of correlation becomes significant only for a higher utilization. Overall the effects of correlations are smaller for this queueing system than for the previous one where inter-arrival and service times are correlated. 

\section{Conclusions}

The paper introduces phase type representations of exponential distributions that allow one to model correlated exponential distributions by starting the distribution in some joint states or by routing the output of one distribution to the next one by choosing the initial state of the second distribution depending on the exit state of the first distribution. The proposed results extend previously published phase type representations for exponential distributions. The representations of the exponential distribution can be  used a building blocks for phase type distributions and Markovian arrival processes that can be integrated in stochastic models like queues with correlated arrival and service times or fault trees with correlated failure times. This expansion has been shown for hyperexponential and Erlang distributions. 

From a theoretical perspective it is still open whether the phase type representation for exponential distributions  that has been derived for positive correlation is optimal in the sense that no representation with less states resulting in the same correlation exists. For negative correlation it has been shown that phase type representations exist that result in a smaller coefficient of correlation with the same number of states but it seems hard to find those representations for phase type distributions of a larger dimension. Another open question is whether more compact phase type representations can be found that are usable as building blocks for Markovian arrival processes. Apart form the applications shown in the paper, correlated phase type distributions can be integrated in many other stochastic models. 

\bibliographystyle{plain}

\begin{thebibliography}{10}

\bibitem{Aa95}
O.~O. Aalen.
\newblock Phase type distributions in survival analysis.
\newblock {\em Scand. J. Statist.}, 2:447--463, 1995.

\bibitem{ArGH10}
J.~R. Artalejo, A.~Gomez-Corral, and Q.~M. He.
\newblock Markovian arrivals in stochastic modelling: a survey and some new
  results.
\newblock {\em SORT: Statistics and Operations Research Transactions},
  34(2):101--156, 2010.

\bibitem{BaLa09}
N.~Balakrishnan and Chin-Diew Lai.
\newblock {\em Continuous Bivariate Distributions}.
\newblock Springer, 2 edition, 2009.

\bibitem{BMSPH12}
Dario Bini, Beatrice Meini, S.~Steff{\'{e}}, Juan~F. P{\'{e}}rez, and Benny~Van
  Houdt.
\newblock {SMCS}olver and {Q-MAM:} tools for matrix-analytic methods.
\newblock {\em {SIGMETRICS} Perform. Evaluation Rev.}, 39(4):46, 2012.

\bibitem{BlNi10}
Mogens Bladt and Bo~Friis Nielsen.
\newblock On the construction of bivariate exponential distributions with an
  arbitrary correlation coefficient.
\newblock {\em Stochastic Models}, 26(3):295--308, 2010.

\bibitem{BBS20}
Andreas Blume, Peter Buchholz, and Clara Scherbaum.
\newblock Markovian arrival processes in multi-dimensions.
\newblock In Marco Gribaudo, David~N. Jansen, and Anne Remke, editors, {\em
  Quantitative Evaluation of Systems - 17th International Conference, {QEST}
  2020, Vienna, Austria, August 31 - September 3, 2020, Proceedings}, volume
  12289 of {\em Lecture Notes in Computer Science}, pages 175--192. Springer,
  2020.

\bibitem{BK17}
Peter Buchholz and Jan Kriege.
\newblock Fitting correlated arrival and service times and related queueing
  performance.
\newblock {\em Queueing Syst. Theory Appl.}, 85(3-4):337--359, 2017.

\bibitem{BKF14}
Peter Buchholz, Jan Kriege, and Iryna Felko.
\newblock {\em {Input Modeling with Phase-Type Distributions and Markov Models
  - Theory and Applications}}.
\newblock Springer, 2014.

\bibitem{CiBS21}
Ismail Civelek, Bahar Biller, and Alan Scheller{-}Wolf.
\newblock Impact of dependence on single-server queueing systems.
\newblock {\em European Journal of Operational Research}, 290(3):1031--1045,
  2021.

\bibitem{Cuma82}
A.~Cumani.
\newblock On the canonical representation of homogeneous {M}arkov processes
  modeling failure-time distributions.
\newblock {\em Micorelectronics and Reliability}, 22(3):583--602, 1982.

\bibitem{FeWh98}
A.~Feldmann and W.~Whitt.
\newblock Fitting mixtures of exponentials to long-tail distributions to
  analyze network performance models.
\newblock {\em Performance Evaluation}, 31:245--279, 1998.

\bibitem{GMSR18}
Bel{\'{e}}n Garc{\'{\i}}a{-}Mora, Cristina Santamar{\'{\i}}a, and Gregorio
  Rubio.
\newblock Modeling dependence in the inter-failure times. an analysis in
  reliability models by markovian arrival processes.
\newblock {\em J. Comput. Appl. Math.}, 343:762--770, 2018.

\bibitem{He12}
Qi-Ming He.
\newblock Analysis of a continuous time {SM[K]/PH[K]/1/FCFS} queue: age
  process, sojourn times, and queue lengths.
\newblock {\em J. Syst. Sci. Complex}, 25(1):133--155, 2012.

\bibitem{HeZV12}
Qi-Ming He, Hanqin Zhang, and Juan Vera.
\newblock On some properties of bivariate exponential distributions.
\newblock {\em Stochastic Models}, 28(2):187--206, 2012.

\bibitem{HeML03}
Armin Heindl, Kenneth Mitchell, and Appie van~de Liefvoort.
\newblock The correlation region of second-order maps with application to
  queueing network decomposition.
\newblock In Peter Kemper and William~H. Sanders, editors, {\em Computer
  Performance Evaluations, Modelling Techniques and Tools. 13th International
  Conference, {TOOLS} 2003, Urbana, IL, USA, September 2-5, 2003, Proceedings},
  volume 2794 of {\em Lecture Notes in Computer Science}, pages 237--254.
  Springer, 2003.

\bibitem{HoTB05}
G.~Horv{\'a}th, M.~Telek, and P.~Buchholz.
\newblock A {MAP} fitting approach with independent approximation of the
  inter-arrival time distribution and the lag-correlation.
\newblock In {\em QEST}, pages 124--133. IEEE CS Press, 2005.

\bibitem{Houdt12}
Benny~Van Houdt.
\newblock A matrix geometric representation for the queue length distribution
  of multitype semi-markovian queues.
\newblock {\em Perform. Evaluation}, 69(7-8):299--314, 2012.

\bibitem{Ke79}
Frank~P. Kelly.
\newblock {\em {Reversibility and Stochastic Networks}}.
\newblock Cambridge University Press, 1979.

\bibitem{KeS69}
J.~G. Kemeny and J.~L. Snell.
\newblock {\em Finite Markov chains}.
\newblock University series in undergraduate mathematics. VanNostrand, New
  York, repr edition, 1969.

\bibitem{Ku89}
Vidyadhar~G. Kulkarni.
\newblock A new class of multivariate phase type distributions.
\newblock {\em Oper. Res.}, 37(1):151--158, 1989.

\bibitem{LHB06}
Joke Lambert, Benny van Houdt, and Chris Blondia.
\newblock Queue with correlated service and inter-arrival times and its
  application to optical buffers.
\newblock {\em Stochastic Models}, 22(2):233--251, 2006.

\bibitem{LiPF14}
Yang{-}Kuei Lin, Michele~E. Pfund, and John~W. Fowler.
\newblock Processing time generation schemes for parallel machine scheduling
  problems with various correlation structures.
\newblock {\em J. Sched.}, 17(6):569--586, 2014.

\bibitem{Luca91}
D.~M. Lucantoni.
\newblock New results on the single server queue with a batch {M}arkovian
  arrival process.
\newblock {\em Stochastic Models}, 7(1):1--46, 1991.

\bibitem{Neut79}
M.~F. Neuts.
\newblock A versatile {M}arkovian point process.
\newblock {\em Journal of Applied Probability}, 16:764--779, 1979.

\bibitem{Neut81}
M.~F. Neuts.
\newblock {\em Matrix-geometric solutions in stochastic models}.
\newblock Johns Hopkins University Press, 1981.

\bibitem{RiDS04}
A.~Riska, V.~Diev, and E.~Smirni.
\newblock An {EM}-based technique for approximating long-tailed data sets with
  {PH} distributions.
\newblock {\em Perform. Eval.}, 55:147--164, 2004.

\bibitem{Stew09}
W.~J. Stewart.
\newblock {\em Probability, Markov Chains, Queues, and Simulation}.
\newblock Princeton University Press, 2009.

\end{thebibliography}

\begin{appendix}

\section{Proof of Theorem~\ref{th:first_phase}}

Let $\iniv_t = \iniv e^{t \mx{D}}$ the transient vector of the APHD. Since the APHD describes an exponential distribution with rate $1$ we have $\iniv_t\idvt = e^{-t}$ and $\iniv_t(n)\mu_n = e^{-t}$. This implies $\iniv_0(n)\mu_n = 1\Rightarrow \iniv_0(n)=1/\mu_n$ where $\iniv_0 = \iniv$. The values of $\iniv_t(i)$ $(i=2,\ldots,n-1$) are given by the following differential equations
$$
\frac{d\iniv_t(i)}{dt} = -\mu_i\iniv_t(i) + \mu_{i-1}\iniv_t(i-1)
$$
with initial conditions $\iniv_0(i) = \iniv(i)$. Since $\frac{d\iniv_t(n)}{dt}=-e^{-t}/\mu_n \wedge \iniv_t(n)\mu_n = e^{-t} \Rightarrow \iniv_t(n-1)\mu_{n-1} = (1-1/\mu_n)e^{-t} \Rightarrow \iniv(n-1)=1/\mu_{n-1}(1-1/\mu_n)$. Now the representation for $\iniv_t(n-1)$ can be substituted in the next differential equation resulting in $\iniv_t(i) = 1/\mu_i\prod_{j=i+1}^n\left(1-1/\mu_j\right)$ and 
$$
\iniv_t(1) = e^{-t}\left(1- \sum_{i=2}^n \frac{1}{\mu_i}\prod_{j=i+1}^n\left(1-\frac{1}{\mu_j}\right)\right)=e^{-t}\left(\prod_{j=2}^n\left(1-\frac{1}{\mu_j}\right) \right) \Rightarrow \iniv_t(1) = \iniv(1)e^{-t}
$$
which implies $\mu_1 = 1$.

\section{Proof of Theorem~\ref{th:transform}}

We build the reversed process which has according to \cite[Theorem 1.12]{Ke79} the same equilibrium distribution as the original process. This implies that the flow from $i$ to $j$ in the original process equals the flow in the reverse direction in the modified process. This implies that every path from the initial state to the absorbing state corresponds to an identical path that starts in the absorbing state and ends in the initial state in the original process. The identity of the vectors follows from the identity of the reversed paths.  

\section{Proof of Theorem~\ref{th:identity}}

Since $\exitv = (0,\ldots,0, 1)$ we have $\inivp = (0,\ldots,0, 1)$ which is the required initial vector after reversing the order of the states. Furthermore, we have
$$
\mu'_{i,i-1} = \mu_{i-1}\frac{\iniv(i-1)}{\iniv(i)} = \mu_{i-1}\frac{1}{\mu_{i-1}}\left(1-\frac{1}{\mu_i}\right)/\frac{1}{\mu_i} = \mu_i - 1
$$
because $\mu_{i-1,i} = \mu_{i-1}$ for $i=1,\ldots,n$. 

\section{Proof of Theorem~\ref{th:stepwise}}

 Let $\mx{g} = \mx{d}-\mx{f}\ge \mx{0}$ and $\inivp_t=(\addv_t, p_t)$ where $\addv_t$ is of length $n$, which equals  the vector at time $t$ for $(\iniv, \mx{D})$,  and $p_t$ is a scalar. 
Since $\iniv_t\mx{d} = \iniv_t(\mx{f}+\mx{g})=e^{-t}$, we have $p_t = p e^{-t}$ and $p_t \mu = p \mu e^{-t} \Rightarrow (1-p)\iniv_t\mx{g} = (1 - p\mu)e^{-t}$. This can be achieved by choosing $\mx{g} = q\mx{d}$ which implies $\mx{f} = (1-q)\mx{d}$. Thus
$$
(1-p)q\iniv_t\mx{d} = (1-p)qe^{-t}= (1-p\mu)e^{-t} \Leftrightarrow
\mu = \frac{1-q+pq}{p}
$$

\section{Proof of Theorem~\ref{th:map}}

Consider an APHD of order $2$ that represents a normalized exponential distribution and let
$$
\iniv = (\iniv(1), \iniv(2)), \ \mx{D} = \left(\begin{array}{cc} -\mu_1 & \mu_{1,2}\\ 0 & -\mu_2\end{array}\right), 
\mx{d} = \left(\begin{array}{c} \mu_1-\mu_{1,2}\\\mu_2\end{array}\right), \iniv_t = \statv e^{\mx{d}t} .
$$
Since the distribution describes an exponential distribution with rate $1$ we have $\iniv_t\idvt = e^{-t}$ and $\iniv_t\mx{d} = e^{-t}$. We assume that $\statv > \mx{0}$ because otherwise one phase can be dropped and we end up with an exponential distribution with a single phase. Furthermore, $\mu_1, \mu_2 \ge 1$ because otherwise $\iniv_t \idvt = e^{-t}$ could not hold. If $\mu_1 = 1$, then Theorem~\ref{th:stepwise} applies and according to (\ref{eq:item3})  $\mx{a}(2) =1$ holds which implies $\mx{a}(1) =1$ or $\exitv(1) = 0$. In both cases the representation is not output flexible and the minimum or maximum of (\ref{eq:map_lp}) becomes $0$.  For $\mu_1 > 1$ $\lim_{t \rightarrow \infty} \iniv_t(1)/\iniv_t(2) = 0$ and $\lim_{t\rightarrow \infty}\iniv_t(2) = \lim_{t \rightarrow \infty} e^{-t}$ which implies $\mu_2=1$ because $\iniv_t(2)\mu_2=e^{-t}$. Additionally, $\statv(1)(\mu_1-\mu_{1,2})+1-\statv(1)=1\Rightarrow \statv(1)(\mu_1-\mu_{1,2}-1)=0\Rightarrow \mu_1-\mu_{1,2} = 1$ which results in the second canonical form with $\mx{m}=(1,1)^T$. Again the minimum and maximum of (\ref{eq:map_lp}) becomes $0$.

\section{Proof of Theorem~\ref{th:map_moments}}

Let $\iniv(i,j)$ the probability of starting two distributions $(\iniv,\mx{D})$ to achieve $\rho^\pm_{(\iniv,\mx{D})}$. Then
$$
\rho^\pm_{(\iniv,\mx{D})} = \sum_{i=1}^n\sum_{j=1}^n \iniv(i,j)\mx{m}(i)\mx{m}(j) .
$$
By choosing $\intv(f_i,e_j)=\intv(n-i+1,n-j+1)$ we have 
$$
\sum_{i=1}^n\sum_{j=1}^n \intv(f_i,e_j)\mx{a'}(f_i)\mx{m'}(e_j) =  \sum_{i=1}^n\sum_{j=1}^n \iniv(i,j)\mx{m}(i)\mx{m}(j) = \rho^\pm_{(\iniv,\mx{D})} .
$$
A larger/smaller coefficient of autocorrelation cannot be generated because this would imply that a larger/smaller coefficient of correlation $\rho_{(\iniv,\mx{D})}$ could be achieved which is not possible since $\rho^\pm_{(\iniv,\mx{D})}$ is the minimum/maximum.

\section{Proof of Theorem~\ref{th:hyperexp}}

For an exponential distributed random variable $X$ with rate $\mu$, $\rho^{+(n)}=\frac{E(XX)-\mu^{-2}}{\mu^{-2}}$ where $E(XX)$ is the expectation of two successive occurrences of $X$ and $\rho^{+(n)}$ is computed according to (\ref{eq:opt_rho}). Observe that $E(XX)= \rho^{+(n)}\mu^{-2}+\mu^{-2}$ and  for the exponential distribution $\sigma^2_X= E(X)^2 = \mu^{-2}$ and $E(X^2) = 2E(X)^2$. If we plug this into (\ref{eq:cc_expanded}), we obtain for the maximal coefficient of correlation of the hyperexponential distribution with $n$ phases where the $i$th phase has been substituted by an APHD of order $n_i$ in the first canonical form.
$$
\frac{\sum\limits_{i=1}^n\iniv(i)\sum\limits_{j=1}^{n_i}\iniv_i(j)(\mx{m}_i(j))^2-E(T)^2}{\sigma^2} = 
\frac{\sum\limits_{i=1}^n \iniv(i)\frac{\rho^{+(n_i)}+1}{\mu_i^2}-E(T)^2}{E(T^2)-E(T)^2} = 
\frac{\sum\limits_{i=1}^n \iniv(i)\frac{\rho^{+(n_i)}}{\mu_i^2}+\frac{1}{2}E(T^2)-E(T)^2}{E(T^2)-E(T)^2} .
$$
The limit follows because $\lim_{n_i \rightarrow \infty} \rho^{+(n_i)} = 1$.

\section{Proof of (\ref{eq:item3})}

To compute $\exitvp$, one first has to observe that every state $i=1,...,n$ is entered with probability $\iniv(i)$ multiplied with $(1-p)$. Thus, the probability of leaving state $i$ toward $n+1$ or the absorbing state is $(1-p)\exitv(i)$ and if $i$ is left, the absorbing state is entered with probability $q$. The last state is entered with probability $p$ as the first state and with probability $(1-p)(1-q)$ after some other non-absorbing state. The sum of both values gives the required result.

If the process starts in state $i$ ($1 \le i \le n$), then the time to leave the first $n$ states equals $\mx{m}(i)$ because this part of the matrix is not modified. Afterwards with probability $(1-q)$ state $n+1$ is entered and the sojourn time in this state equals $1/\mu = p/(1-q+pq)$. The sum of both values gives the required result. For state $n+1$ the sojourn time equals $1/ \mu$.


The values for $\mx{a}(i)$ remain if the distribution is left from one of the first $n$ states because the sub-matrix $\mx{D}$ remains unchanged.  If the distribution is left from the last state, then the sojourn time in the last state equals $1/\mu=p/(1-q+pq)$ and with probability $(1-p)(1-q)$ the last state is entered from one of the first states. In this case the sojourn time in the first states is $1$ because $(\iniv,\mx{D})$ represents an exponential distribution with rate $1$. Thus, the unconditioned time equals $p+(1-p)(1-q)=1-p+pq$ which has to be divided by $1-p+pq$ the probability of leaving from the last state which gives $1$.

\section{Proof of (\ref{eq:item4})}

It has to be noted that the numbers of states in the original APHD $(\iniv,\mx{D})$ are incremented by $1$ since a new state is prepended. The equation for $\inivp$ follows immediately from the definition of the vector. To depart from the first state, the distribution has to be entered in the first state and has to leave the first state toward the absorbing state. The former probability is $p$ and the probability of leaving toward the absorbing state equals $1/\mu$, together this gives $p/\mu$. The probability of entering some state $i$ either initially or from the first state equals $\iniv(i-1)$ multiplied with $(1-p)$, if entered initially, and multiplied with $p(\mu-1)/\mu$, if a state is entered from the first state. Thus, entry probabilities are scaled with $(\mu-p)/\mu$ and matrix $\mx{D}$ remains unchanged which implies that the exit probabilities have to be multiplied with the scaling factor. 

Since matrix $\mx{D}$ remains unchanged, the time to enter the absorbing state remains unchanged. If the process begins in the first state, then the sojourn time in this state equals $\mu^{-1}$, afterwards the absorbing state is entered with probability $\mu^{-1}$ and with probability $1-\mu^{-1}$ one of the states $2,\ldots,n$ is entered and the mean sojourn time in these states before entering the absorbing state is $1$. Together this gives
$$
\frac{1}{\mu^2} + \frac{\mu-1}{\mu}\left(1-\frac{1}{\mu}\right) = \frac{1}{\mu^2} = \frac{(\mu-1)(\mu+1)}{\mu^2} = 1 .
$$
If the APHD is left from the first state, the sojourn time was $\mu^{-1}$. If the APHD is left from one of the states $i=2,\ldots,n$, then it has been entered with probability $(1-p)$ in one of the states $i=2,...,n$, the sojourn time equals then $\mx{a}(i-1)$, with probability $p(\mu-1)/\mu$ it has been entered in state $1$ and the sojourn time becomes $\mx{a}(i-1)+\mu^{-1}$. The sum of both conditional sojourn times multiplied with the probabilities and scaled by $\left(1-p/\mu\right)^{-1}$ gives the required result.  

\end{appendix}
\end{document}